\documentclass[aps,twocolumn,prd,reprint,superscriptaddress,showpacs,preprintnumbers,nofootinbib,10pt,floatfix]{revtex4-1}
\RequirePackage[l2tabu, orthodox]{nag}

\usepackage{microtype} 

\usepackage{amsmath}
\usepackage{amstext, amssymb, amsthm, amsfonts, slashed}
\usepackage{mathtools}
\usepackage{mhchem}
\usepackage{graphicx}
\usepackage{booktabs}
\usepackage{slashed}
\usepackage{dcolumn}
\usepackage{bm}
\usepackage{dsfont}
\usepackage[usenames,dvipsnames]{xcolor}
\usepackage[normalem]{ulem}

\usepackage{url}
\usepackage[colorlinks=true,urlcolor=blue,linkcolor=blue,citecolor=blue,hypertexnames]{hyperref}
\usepackage[noabbrev]{cleveref}
\usepackage{braket}
\usepackage{simplewick}
\usepackage{color}

\usepackage{float}

\graphicspath{ {./graphics/} }

\DeclareSymbolFont{symbolsC}{U}{pxsyc}{m}{n}

\def\eq#1{\eqref{#1}}

\newcommand{\p}{\partial}

\newcommand{\mr}[1]{\mathrm{#1}}
\newcommand{\f}[2]{\frac{#1}{#2}}
\newcommand{\s}[1]{\slashed{#1}}

\newcommand{\df}{{\mathrm d}}

\newbox{\ORCIDicon}
\sbox{\ORCIDicon}{\large
                  \includegraphics[width=0.8em]{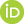}}

\begin{document}
\begin{flushright}
KUNS-2836
\end{flushright}
\vspace{-1cm}
\title{Gravitational instantons and anomalous chiral symmetry breaking}
\author{Yu \surname{Hamada}\,\href{https://orcid.org/0000-0002-0227-5919}
				{\usebox{\ORCIDicon}}
				}
\affiliation{Department of Physics, Kyoto University, Kyoto 606-8502, Japan}
\author{Jan M. \surname{Pawlowski}\,\,\href{https://orcid.org/0000-0003-0003-7180}
                               	{\usebox{\ORCIDicon}}
				}
\affiliation{Institut f{\" u}r Theoretische Physik, Universit{\" a}t Heidelberg, Philosophenweg 16, 69120 Heidelberg, Germany}
\affiliation{ExtreMe Matter Institute EMMI, GSI Helmholtzzentrum f{\" u}r Schwerionenforschung mbH, Planckstr. 1, 64291 Darmstadt, Germany}
\author{Masatoshi \surname{Yamada}\,\href{https://orcid.org/0000-0002-1013-8631}
                               {\usebox{\ORCIDicon}}
                               }
\affiliation{Institut f{\" u}r Theoretische Physik, Universit{\" a}t Heidelberg, Philosophenweg 16, 69120 Heidelberg, Germany}

\begin{abstract}
We study anomalous chiral symmetry breaking in two-flavour QCD induced by gravitational and QCD-instantons within asymptotically safe gravity within the functional renormalisation group approach. Similarly to QCD-instantons, gravitational ones, associated to a K3-surface connected by a wormhole-like throat in flat spacetime, generate contributions to the 't~Hooft coupling  proportional to  $\exp(-1/g_N)$ with the dimensionless Newton coupling $g_N$. Hence, in the asymptotically safe gravity scenario with a non-vanishing fixed point coupling $g_N^*$, the induced 't Hooft coupling is finite at the Planck scale, and its size depends on the chosen UV-completion. Within this scenario the gravitational effects on anomalous $U(1)_A$-breaking at the Planck scale may survive at low energy scales. In turn,  fermion masses of the order of the Planck scale cannot be present. This constrains the allowed asymptotically safe UV-completion of the Gravity-QCD system. We map-out the parameter regime that is compatible with the existence of light fermions in the low-energy regime.
\end{abstract}
\maketitle

\section{Introduction}
One of the long-standing problems in theoretical high-energy physics is the construction of a well-defined UV-completion of particle physics including quantum gravity. Amongst the promising candidates for such a UV-completion is the asymptotic safety scenario~\cite{Weinberg:1976xy, Hawking:1979ig, Weinberg:1980gg, Reuter:1996cp, Souma:1999at}. In asymptotically safe gravity, the theory approaches a non-trivial interacting ultraviolet fixed point (Reuter fixed point) for large momentum scales in contradistinction to the perturbative free Gau\ss ian fixed point. The physics parameters of such a gravity-matter system are its UV-relevant and marginal couplings, including the Newton coupling and cosmological constant, whose dimensionless versions approach finite values. The set of UV-IR trajectories emanating from the UV-fixed point provide us with the set of potential IR-scenarios below the Planck scale for asymptotically safe gravity-matter systems. Evidently, the physics of these IR-completions crucially depends on the details of the running of the matter couplings from the UV fixed point towards below Planck-scale momentum scales. This running is governed by he gravity-induced anomalous dimensions of matter interactions. 

An investigation of asymptotically safe particle physics asks for a non-perturbative treatment and most studies are based on the functional renormalisation group (fRG) approach. The impress progress in the past two decades enables us to access more intricate questions such as gravitational catalysis of strong chiral symmetry breaking discussed in the present paper. For reviews and textbooks on asymptotically safe gravity and gravity-matter systems, see e.g.\  \cite{Niedermaier:2006wt, Litim:2011cp, Ashtekar:2014kba, Percacci:2017fkn, Bonanno:2017pkg, Eichhorn:2018yfc, Wetterich:2019qzx, Reuter:2019byg, Pereira:2019dbn, Reichert:2020mja, Pawlowski:2020qer}. 

The non-trivial UV-dynamics of asymptotically safe gravity-matter systems opens the possibility for the intriguing possibility of gravitational catalysis of chiral symmetry breaking. If present, it first of all provides a selection criterion for viable systems as the natural mass scale of gravity-induced chiral symmetry breaking is the Planck scale. In turn, it may open the door to signatures  for asymptotic safety far below the Planck scale, being accessible at the LHC and beyond, e.g.\ the FFC. For these reasons gravitational catalysis of chiral symmetry breaking has been investigated in a series of works in (Euclidean) flat spacetimes, see e.g.\ \cite{Eichhorn:2011pc,Eichhorn:2011ec, Eichhorn:2016vvy, Meibohm:2016mkp, Eichhorn:2017eht}. There it has been shown, that rather generic four-fermi systems do not show gravity-induced chiral symmetry breaking for all flavour numbers. Background-curvature induced chiral symmetry breaking has been observed in \cite{Ebert:2008pc}, see also  \cite{Buchbinder:1989fz,Buchbinder:1989ah,Inagaki:1993ya,Sachs:1993ss,Elizalde:1995kg,Kanemura:1995sx,Inagaki:1995bk,Inagaki:1996nb,Geyer:1996wg,Geyer:1996kg,Miele:1996rp,Vitale:1998wm,Inagaki:1997kz,Hashida:1999wb,Gorbar:1999wa,Gorbar:2007kd,Hayashi:2008bm,Gorbar:2008sp,Inagaki:2010py,Sasagawa:2012mn,Gies:2013dca, Flachi:2014jra, Flachi:2015sva, Dvali:2017mpy}. For curvature bounds in asymptotically safe gravity-matter systems see  \cite{Gies:2018jnv}. 

In any case, we expect sizable topology-changing fluctuations in the UV, which can be taken into account within an expansion about a background spacetime with non-trivial topology. Such topology-changing processes are induced by gravitational instantons  \cite{Eguchi:1977iu,Eguchi:1978xp,Eguchi:1979yx,Hawking:1976jb, Hebecker:2019vyf} in the semi-classical limit. Similarly to QCD-instantons, they induce (anomalous) $U(1)_A$-violating fermionic self-interactions. In particular they generate contributions to the coupling of the 't~Hooft term introduced in QCD in  \cite{Kobayashi:1971qz,tHooft:1976rip,tHooft:1976snw}. Due to the UV fixed-point scaling the contributions to the 't~Hooft coupling is sizable but finite in the UV fixed point regime while it decouples quickly in the IR-regime with a classical running of the gravity coupling: roughly speaking, $M_\mathrm{pl}^{-1}$ is an effective IR-cutoff for the size of gravitational instantons. In addition, the size of the gravitational instanton is a quasi-zero mode when the instanton is much smaller than $M_\mathrm{pl}^{-1}$ because of the (quantum) scale invariance of the theory. Therefore the effect is finite even at an arbitrarily high momentum scale. 

In the present work we put forward a novel mechanism for dynamical breaking of the chiral symmetry. It is triggered by gravitational instantons in asymptotically safe gravity-matter systems. Specifically, we consider two-flavour massless QCD coupled to gravity, and investigate the RG flow of two four-fermion interactions in a two-channel approximation: we consider the 't~Hooft vertex and the scalar--pseudo-scalar channel. Divergences at a RG-scale in the flow of the latter channel indicate resonant interactions and hence signal spontaneous chiral symmetry breaking. The flow of the 't~Hooft vertex in QCD has been evaluated in \cite{Pawlowski:1996ch}. There it has been shown that the semi-perturbative flow of the 't Hooft coupling together with an (semi-classical) initial condition in the UV gives rise to the right amount of anomalous chiral symmetry breaking at cutoff scales of about 1\,GeV. 

In the present work we take a first step towards the evaluation of gravity-induced $U(1)_A$ breaking within a  phenomenological approach similar to the instanton-liquid in QCD. We derive the RG-equations in the presence of gravitational instantons, and consider the prefactors of the respective topological terms as phenomenological input parameters. This allows us to evaluate not only one-instanton contributions to the flow, but also an interacting gravitational instanton-liquid within a phenomenological approach. Such an instanton-liquid may well be present in the strongly-correlated UV-sector close to the Reuter fixed point. 

Within this setup we analyse the subset of the parameter space for which chiral symmetry breaking is triggered by gravitational instantons. The necessity of the occurrence of light (infrared) fermions allows us to  exclude a quite wide range of the parameter space. In turn, part of the parameter space is compatible with the observation of Standard model fermions but may lead to signatures beyond the currently accessible energy range. 

The paper is organised as follows: In Sec.~\ref{sec:QCD-gravity} we set up the two-flavour Gravity-QCD system investigated in the present work. In Secs.~\ref{sec:GravInst} and \ref{sec:tHooft} we briefly review  gravitational instantons, and anomalous chiral symmetry breaking due to gravitational instantons. In Sec.~\ref{sec:GravCat} we compute numerically the UV-IR flows in the Gravity-QCD system in the presence of gravitational instantons, and investigate the respective chiral symmetry breaking. Sec.~\ref{sec:Discussion} contains discussions of the results of the present work, and conclusions.

\section{QCD coupled to gravity}
\label{sec:QCD-gravity}
\begin{figure*}[tbp]
 \centering
 \includegraphics[width=0.45\textwidth]{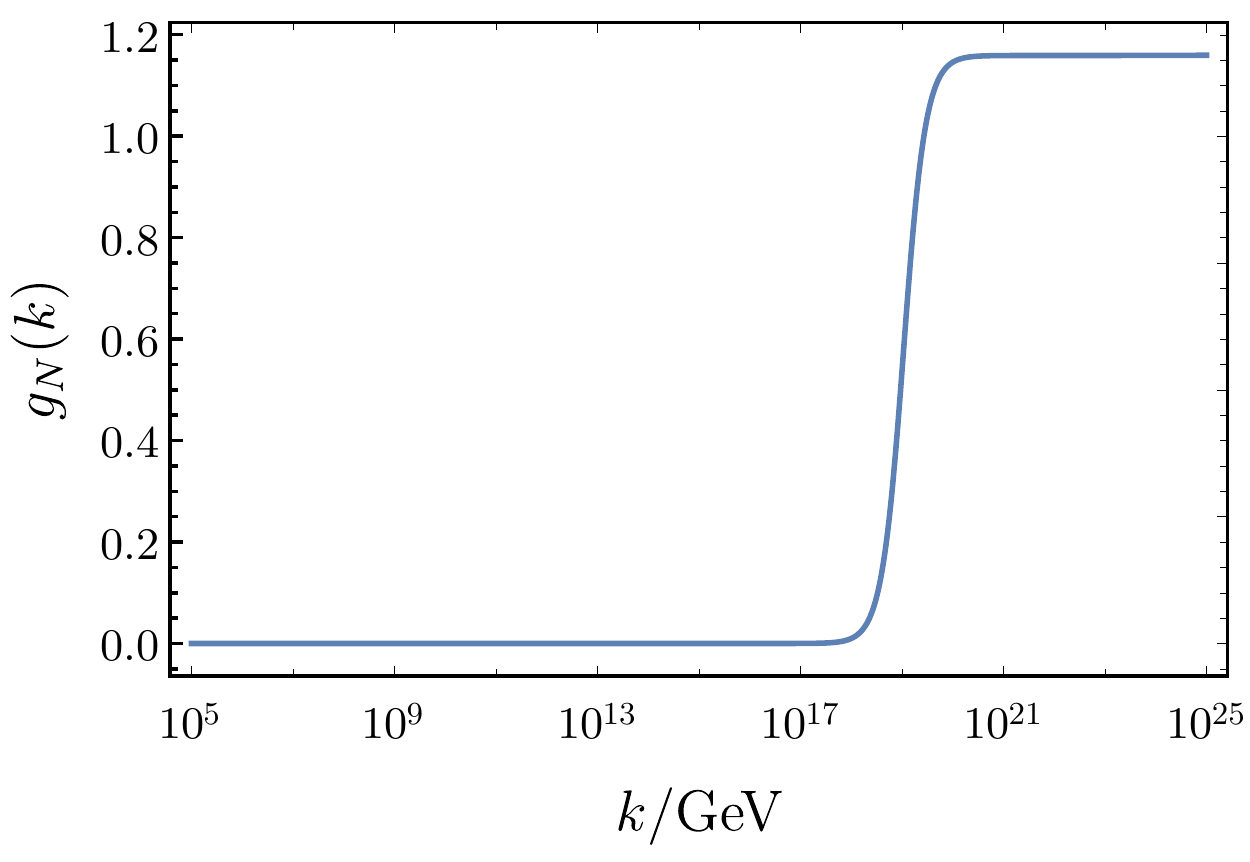}  
\hspace{0.05\textwidth}
 \includegraphics[width=0.45\textwidth]{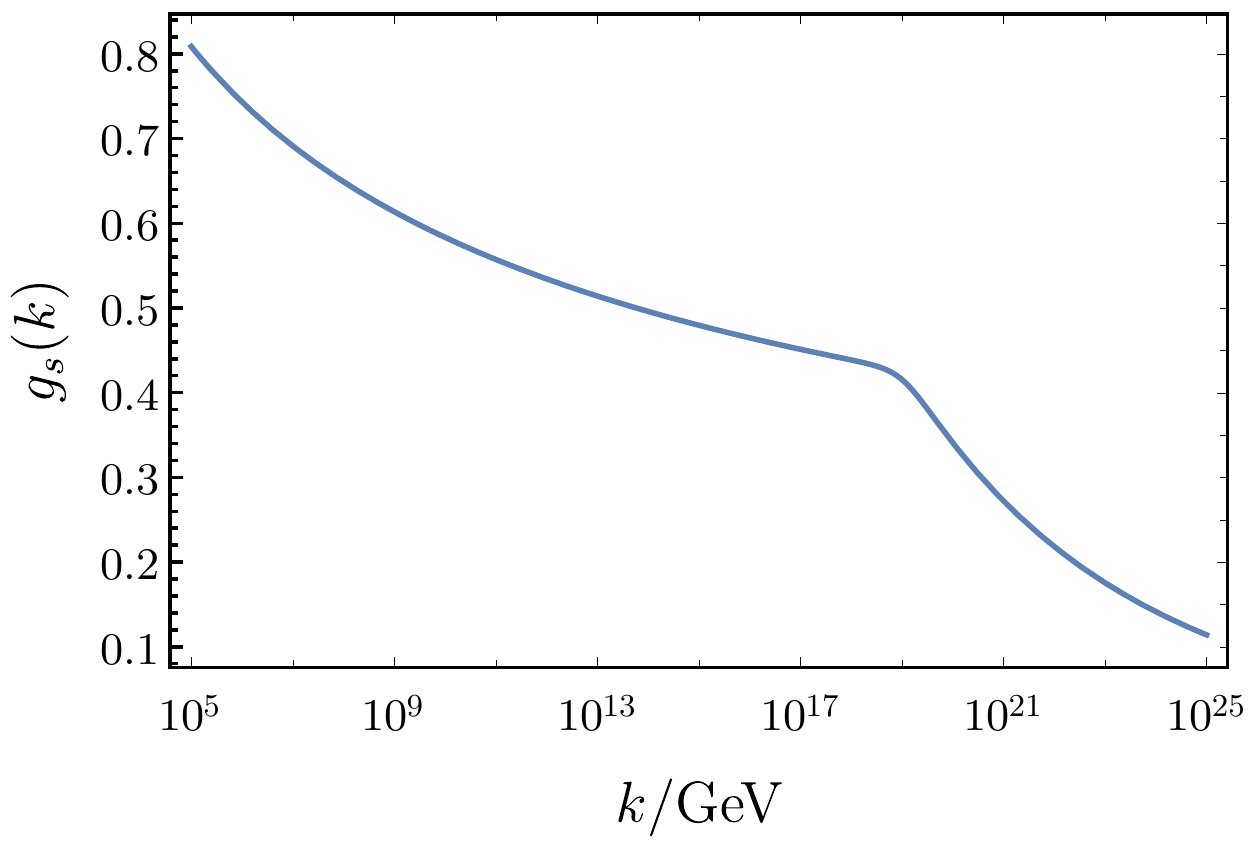}
 \caption{IR-UV RG-flow of $g_N(k)$ and $g_s(k)$. The flow is initiated close to the electroweak scale at the initial scale  $k_\textrm{IR}=10^2~\textrm{GeV}$, and the initial conditions for $g_N$ and $\alpha_s=g_s^2/(4 \pi)$ are the physical ones, $g_N= 6.71 \times 10^{-35 }$, and $\alpha_s=0.118$. The flow of $g_N$ approaches the UV-fixed point value $g^*_{N}\simeq 1.15907$ and $\alpha_s$ tends towards zero. Below the Planck scale, $M_\textrm{pl}\simeq 1.22 \times 10^{19 }$\,GeV, the Newton coupling has a classical running, $g_N\propto k^2$, above the Planck scale it is rapidly approaching the fixed point value. The strong coupling runs perturbatively below the Planck scale and decays rapidly due to the gravity corrections above the Planck scale. }
 \label{fig:gs-gNFlow}
\end{figure*}

In this paper, we consider massless two-flavour QCD coupled to gravity. Its momentum-cutoff scale dependence is investigated with the functional renormalisation group (fRG) approach.  The flow of correlation functions and couplings is derived from the master equation for the scale-dependent (one-particle irreducible) gauge-fixed effective action  $\Gamma_k$, where $k$ is the infrared cutoff scale: momenta $p^2 \lesssim k^2$ are suppressed, and hence $\Gamma_k$ agrees at $k=0$ with the full quantum 1PI effective action $\Gamma$. 

Both, the required gauge-fixing and the infrared regulator  gravity necessitate the introduction of a background metric and a respective split of the full metric, for a detailed discussion see e.g.\ \cite{Pawlowski:2020qer}. Here we consider a standard linear split, 
\begin{align}\label{eq:Split}
g_{\mu\nu} = \bar g_{\mu\nu} +  \sqrt{Z_h G_N} h_{\mu\nu}\,, \qquad A_\mu =\bar A_\mu +a_\mu\,, 
\end{align}
where the dimension one fluctuation field $h_{\mu\nu}$ carries the gravity quantum dynamics. The background split for the gluon is introduced for convenience as it allows for an expansion of the scale-dependent quantum effective action $\Gamma_k$ about topologically non-trivial configurations, see \cite{Pawlowski:1996ch}. Apart from the graviton and gauge fields we also have the matter fields, the two-flavour quarks $q,\bar q$. Together with the auxiliary ghost fields introduced within the Faddeev-Popov quantisation of gravity and QCD, the total field content of gravity coupled to $N_f=2$-flavour QCD is given by the backgrounds $(\bar g_{\mu\nu}, \bar A_\mu)$ and the dynamical fluctuation fields 
\begin{align}
\phi= (\phi_\textrm{grav},\phi_\textrm{mat})\,,
\end{align}
with
\begin{align}
\phi_\textrm{grav}=(h_{\mu\nu}, c_\mu, \bar c_\mu)\,,\qquad \phi_\textrm{mat}=(A_\mu, c,\bar c, q,\bar q)\,.
\end{align}
The field $\phi$ is the fluctuation super field with the gravity part $\phi_\textrm{grav}$ and the matter part is $\phi_\textrm{mat}$. 

The scale dependence of $\Gamma_k[\bar g, \bar A, \phi]$ is described by the Wetterich equation~\cite{Wetterich:1992yh, Ellwanger:1993mw, Morris:1993qb}, for gravity see \cite{Reuter:1996cp}, 
\begin{equation}\label{eq:Wett}
 \partial_t \Gamma_k = \frac{1}{2} \mathrm{Tr}\left[\frac{1}{\Gamma_k^{(2)}+R_k}\partial_t R_k\right]\,,\quad \textrm{with}\quad t=\log \frac{k}{k_\textrm{ref}}\,, 
\end{equation}
where $k_\textrm{ref}$ is a reference scale and $t$ is (minus) the RG-time. The regulator $R_k(p^2)$ suppresses the infrared propagation of modes $p^2\lesssim k^2$ and vanishes for the ultraviolet modes. This regulator function ins multiplied with the full field-dependent propagator $[\Gamma_k^{(2)}+R_k]^{-1}$, which is matrix-valued in field space. Here, $\Gamma_k^{(2)}$ is the second functional derivative of $\Gamma_k$ with respect to fluctuation fields $\phi$. The operator trace in \eq{eq:Wett} denotes the summation over discrete and integration of continuous variables such as momenta and flavour.  For Grassmann-valued fields the trace involves a minus-sign. For reviews on the FRG and the Wetterich equation, see  Refs.~\cite{Aoki:2000wm,Berges:2000ew,Polonyi:2001se,Pawlowski:2005xe,Gies:2006wv,Delamotte:2007pf,Rosten:2010vm,Braun:2011pp,Reichert:2020mja,Dupuis:2020fhh}.

For interacting theories such as the present Gravity-QCD system, \eq{eq:Wett} has to be solved within a suitable truncation of the full effective action. The aim of the current work is the analysis of a novel mechanism for chiral symmetry breaking triggered by asymptotically safe instantons. This analysis is possible and well-accessible within a simple truncation in Euclidean spacetime, a more quantitative analysis is deferred to future work. The full effective action can be split into its different sectors, 
\begin{align}
\Gamma_k &=\Gamma_\text{grav}+ \Gamma_\text{glue} +\Gamma_\text{mat}+ \Gamma_\textrm{gh}+S_\textrm{gf}\,.  
\label{eq:EffAction} \end{align}
In \eq{eq:EffAction}, $\Gamma_\text{grav}$ stands for the effective action part of the pure graviton sector, obtained with vanishing matter and gauge fields as well as vanishing auxiliary ghost fields. Similarly, the pure glue sector $\Gamma_\text{glue}$ only contains the gluonic gauge field and the graviton, while the matter part $\Gamma_\text{mat}$ vanishes for $q\bar q=0$. Finally, $S_\textrm{gf}$ carries the gauge-fixing terms for QCD and gravity, while $\Gamma_\textrm{gh}$ is the auxiliary ghost sector.   

Here we consider the following qualitative truncation of the full effective action in \eq{eq:EffAction}: In  $\Gamma_\text{grav}$ we consider the full fluctuating two-point function with a wave function renormalisation $Z_h$ and a graviton mass parameter $\mu_h=-2 \Lambda_2$. Furthermore we consider higher correlation functions as derived from the Einstein-Hilbert action with running Newton constant $G_N$ and cosmological constant $\Lambda$: the flows are computed from the fluctuation three-point function with  $G_N=G_3$ and $\Lambda=\Lambda_3$ and all higher couplings are identified with that of the three-point function: $G_{n>3} =G_3$ and $\Lambda_n=\Lambda$. For more details see e.g.\ the recent review \cite{Reichert:2020mja} and references therein. 

For the pure glue part we use a similar, even simpler approximation: $\Gamma_\textrm{YM}$ has the form of the classical Yang-Mills action with running gauge coupling $g_s$. This truncation can be summarised as follows, 
\begin{align}\nonumber 
&\Gamma_\text{grav}^{(2)}= Z_h\, \left[\frac{1}{16\pi G_N}\int \df^4 x\sqrt{g} \left(-\mu_h - R \right)\right]^{(2)} \,, \\[1ex]\nonumber 
&\Gamma_\text{grav}^{(n>3)}= \left[\frac{1}{16\pi G_N}\int \df^4 x\sqrt{g} \left(2\Lambda - R \right)\right]^{(n)}\,, \\[2ex]
&\Gamma_\text{glue} = \frac{1}{4 g_s^2}\int \df^4 x\sqrt{g} \, g^{\mu\rho} g^{\nu\sigma} F^a_{\mu\nu}F^a_{\rho\sigma}\,,
\label{eq:EH-YM}\end{align}
where the superscript ${}^{(n)}$ denotes $n$-derivatives w.r.t.\ the fluctuation field $h$. This entails that $\Gamma^{(2)}_\textrm{grav}$ does not depend on $G_N$. 
In \eqref{eq:EH-YM}, $\sqrt{g}$ denotes the squared determinant of the metric; $R$ is the scalar curvature; $F_{\mu\nu}$ is field strength of the gauge field $A_\mu$. In pure gravity and for the current Gravity-QCD system the $\Lambda_{n>2}$ are rather small and can be safely put to zero, that is, $\Lambda=0$ in Eq.~\eqref{eq:EH-YM}. The mass parameter of the fluctuating graviton is $\mu^*_h \approx -1/2$. While this decreases significantly the fixed point value of the fixed-point Newton coupling, it does not change the qualitative behaviour. Moreover, the flows depend on the anomalous dimension of the fluctuating graviton, $\eta_h = -\partial_t \log Z_h$. Also, the fixed point value $\eta^*_h\approx 1/2$ is positive, and vanishes quickly below the Planck scale. Hence, similarly to $\mu_h$, the graviton anomalous dimension only has a quantitative impact on the present analysis. Note, that this is in contrast to the anomalous dimension of the Newton coupling, that necessarily tends towards $-2$ at the UV fixed point and cannot be dropped. Consequently we also choose $\mu\equiv 0$ for our explicit computations. In summary, the numerical solutions in the current work are computed for $\mu=0$,  $\Lambda=0$ and $\eta_h=0$.  

We use the following combined (background) gauge fixing term 
\begin{align}
S_\text{gf} = \int \df^4x \sqrt{\bar g} \,g^{\mu\nu} \,\left[  \frac{1}{2\xi} (\bar D^{ab}_\mu A^b_\nu)^2 
+ \frac{1}{\alpha}F_\mu F_\nu  \right]\,. 
\end{align}
where $\bar D=\partial - i \bar A_\mu$ is the background-covariant derivative. The gravity gauge fixing $F_\mu$ is given by  
\begin{align}
F_\mu&= \bar g^{\mu\nu} \bar \nabla_\nu h_{\mu\nu} - \frac{1+\beta}{4}\bar \nabla_\mu h\,,  
\end{align}
with the background-metric covariant derivative $\bar\nabla$ and the trace mode $h=\bar g^{\mu\nu} h_{\mu\nu}$. The ghost-dependent part of the effective action is approximated by its classical counterpart, 
\begin{align}
\Gamma_\text{gh} = \int \df^4x \sqrt{\bar g}\, g^{\mu\nu} \left( \bar c \,\p_\mu D_\mu c 
- \bar c_\mu M^{\mu\nu} c_\nu\right) \,,
\end{align}
with 
\begin{align}
M_{\mu\nu}&= \bar g^{\mu\nu}\bar \nabla^2+\frac{1-\beta}{2}\bar \nabla^\mu \bar \nabla^\nu +\bar R^{\mu\nu}\,.  \nonumber 
\end{align}
It is left to specify the matter part. We resort to a combination of the classical Dirac action with running couplings and a two-channel approximation (scalar-pseudoscalar and $U(1)_A$-breaking channel) of the Fierz-complete four-quark interactions. Then the matter part of the effective action reads
\begin{align}\nonumber 
\Gamma_\textrm{mat} =& \, \int \df^4x |e| \Bigl[ \bar{q} \,i\slashed{\nabla}\, q+ \lambda_q \, \left[(\bar{q} q)^2 - (\bar{q } \gamma_5 \tau_a q)^2\right] \\[1ex] 
&\, \hspace{1.5cm}+ 2 \lambda_\mathrm{top} \,\mathrm{det}\, \bar{q}(1-\gamma_5) q+\text{h.c.}\Bigr]\,,
\label{eq:Gferm} \end{align}
with the Paul matrices $\tau_a$ ($a=1,2,3$). In \eq{eq:Gferm} we have suppressed the flavour index: $q\equiv q_i$ ($i=1,2$), and $\nabla_\mu=\p_\mu -i A_\mu-\frac{1}{2}\omega_{\mu ab} J^{ab} $ is the (full) covariant derivative in curved space. Here $\omega_{\mu ab}$ is the spin connection, $J^{ab}=\frac{1}{4}[\gamma^a,\gamma^b]$ is the generator of the Lorentz transformation based on SO(4), and $e^a{}_\mu$ is the vierbein field. This leads us to $\slashed{\nabla}= \gamma^a e_a{}^\mu \nabla_\mu$, and we have used $|e|$ for the  determinant of the vierbein field.  The two four-quark terms in \eq{eq:Gferm} take into account the scalar-pseudoscalar channel (pions and sigma mode) and the axial $U(1)_A$-breaking 't Hooft interaction \cite{Kobayashi:1971qz,tHooft:1976rip,tHooft:1976snw} induced by instantons. 

The approximation \eq{eq:Gferm} does only take into account two channels of the Fierz-complete basis with ten channels in two-flavour QCD. Now we shall argue that this is already sufficient for the present purpose, and indeed constitutes already a semi-quantitative approximation: 

In QCD in flat spacetime, these (and other) four-quark interactions are generated at high scales by quark-gluon box (flow) diagrams proportional to $\alpha_s^2(p)$. Indeed, the flow will generate all four-quark interactions with tensor structures that are compatible with the symmetry of the Dirac action, if the regulator is not breaking these symmetries explicitly. Due to the chiral symmetry of the Dirac term the respective four-quark interactions are invariant under the chiral $SU(2)_R\times SU(2)_L$ symmetry:  
\begin{equation}\label{eq:SU2RL}
q_R \to U_R\, q_R \,, \hspace{2em} q_L \to U_L \,q_L\,,
\end{equation}
where $U_{R/L}\in \textrm{SU(2)}_{R/L}$, and the right- and left-handed quarks are defined with $q_{R/L}= (1\pm \gamma_5)/2 q$. 

While the axial $U(1)_A$-symmetry is a symmetry of the Dirac term, it is broken by the axial anomaly in the quantisation. Accordingly, the resulting four-quark terms are not necessarily invariant under $U(1)_A$-transformations 
\begin{equation}\label{eq:U1A}
q \to e^{i \theta \gamma_5} q\,, \qquad \bar q \to \bar q \, e^{i \theta \gamma_5}\,.
\end{equation}
Indeed, both terms in \eq{eq:Gferm} are invariant under $\textrm{SU(2)}_{R}\times \textrm{SU(2)}_{L}$  transformations, \eq{eq:SU2RL}, but are not invariant under $U(1)_A$-transformations, \eq{eq:U1A}. 

Due to asymptotic freedom the QCD-contributions to the four-quark interactions decay rapidly in the UV. In turn, towards the infrared for momenta $p^2 \lesssim 1-2$\, GeV, the quark-gluon box diagrams increase rapidly with $\alpha_s^2(p)$. Moreover, for $p^2 \lesssim 1$\,GeV the scalar-pseudoscalar channel is getting resonant and the coupling $\lambda_q$ diverges, tantamount to strong spontaneous chiral symmetry breaking. In this regime it is convenient to introduce low energy effective mesonic degrees of freedom, in particular the pion for the scalar-pseudoscalar channel. This infrared dominance of the scalar-pseudoscalar channel is working very efficiently, and while a Fierz-complete basis in two-flavour QCD contains ten tensor structures, only the scalar-pseudoscalar one with the coupling $\lambda_q$ is driving the chiral dynamics. For a quantitative study in quenched and unquenched QCD see \cite{Mitter:2014wpa, Cyrol:2017ewj, Braun:2019aow}, for a recent overview including the relevant literature see \cite{Dupuis:2020fhh}. 

Even though the dynamics is dominantly driven by the scalar-pseudoscalar channel, the axial $U(1)_A$-breaking 't~Hooft interaction \cite{Kobayashi:1971qz,tHooft:1976rip,tHooft:1976snw} is also  important. While it does not drive the chiral dynamics, it is responsible for the anomalously large $\eta$- mass in two-flavour QCD, in 2+1 flavour QCD is triggers the anomalously large $\eta'$-mass via QCD-instanton  effects, see e.g., \cite{tHooft:1986ooh,Hatsuda:1994pi}. Consequently, the two-channel approximation used here is already semi-quantitative, as the contributions of the other eight tensor structures are sub-leading in the vacuum, and can be safely dropped. 

In summary this leaves us with an approximation for the with $N_f=2$ Gravity-QCD system, that is described by the dimensionless couplings 
\begin{align}\label{eq:Couplings}
g_N = G_N k^2\,,\quad g_s\,,\quad \bar \lambda_q = \lambda_q k^2\,,\quad  \bar\lambda_\textrm{top} = \lambda_\textrm{top} k^2\,. 
\end{align}
As discussed above, this is the minimal approximation which suffices to analyse the anomalous gravitational catalysis of chiral symmetry breaking. 

In QCD, the 't~Hooft interaction is generated from non-trivial topological (selfdual) gauge field configurations, the QCD-instantons. In the Gravity-QCD system studied in the present work, we have additional contributions from gravitational topological configurations (gravitational instantons), a brief introduction to the latter is given in the next Sec.~\ref{sec:GravInst}, and the topological contributions both from QCD and gravity are discussed in Sec.~\ref{sec:tHooft}. Here we first restrict ourselves to the Gravity-QCD system in the topologically trivial sector. The flows for the minimal set of dimensionless couplings in \eq{eq:Couplings} is given by that of the dynamical Newton coupling, taken from \cite{Christiansen:2015rva, Meibohm:2015twa, Denz:2016qks, Christiansen:2017cxa}, 
\begin{align}
\partial_t g_N = &\,  2 g_N - \left( \frac{833 }{15} +\frac{133}{30} \left(N_c^2-1\right) +\frac {3599}{600} N_f \right) \frac{g_N^2}{19 \pi} \,,\label{eq:gNFlow}
\end{align}
with the pure gravity part and the contributions from the gluon loops proportional to the number of gluons, $N_c^2-1$, and quark loops proportional to the number of flavours. The flow of the strong coupling receives gravity contributions apart from the universal QCD $\beta$-function, both leading to asymptotic freedom in the UV, taken from \cite{Folkerts:2011jz, Christiansen:2017cxa}, 
\begin{align}
\partial_t g_s= &\, -\left(\frac{11}{3}N_c - \frac{2 N_f}{3} \right) \frac{g_s^3}{16\pi^2}-\frac{g_s g_N}{4 \pi} \,,\label{eq:gsFlow} 
\end{align}
For the flows of the four-quark interactions we restrict ourselves to $N_f=2$. For respective flows in QCD and the Standard model see e.g.\ \cite{Gies:2003dp, Gies:2005as, Braun:2011pp, Mitter:2014wpa, Cyrol:2017ewj, Braun:2019aow}, for gravity contributions see e.g.\ \cite{Eichhorn:2011pc}. In summary this leads us to 
\begin{align}
\partial_t \bar\lambda_q = &\, \left( 2- \frac{N_c^2 - 1}{8\pi^2}\frac{3}{N_c} g_s^2 + \frac{29}{15\pi} g_N \right)  \bar\lambda_q   \nonumber \\[1ex]
& -  \frac{9 N_c^2-24}{256 \pi^2}\frac{3}{N_c} g_s^4 -\frac{5}{16} g_N^2 \nonumber \\[1ex] 
& - \frac{1+2 N_c}{2 \pi^2}\bar\lambda_q^2  -\frac{2(1+ N_c)}{\pi^2} (\bar\lambda_q +\bar\lambda_{\mathrm{top}})\bar\lambda_{\mathrm{top}} \,.\label{eq:qFlow}
\end{align}
The first line provides the dimensional running of $\lambda_q$ including its anomalous deformation due to  mixed $\lambda_q$--gluon and $\lambda_q$--graviton exchange diagram. The second line in \eq{eq:qFlow}  stems from the quark-gluon and quark-graviton exchange diagrams that generate the four-quark interactions  from quark-gravity and quark-gluon interactions. The last line stems from the four-quark self-interactions and is also present in the respective NJL-type four-quark model. 

Finally, the flow of the dimensionless 't Hooft coupling $\bar\lambda_{\textrm{top}}$ takes a form similar to that of $\bar\lambda_q$. In comparison, the gluon and graviton box diagrams and mixed terms are missing, as the Dirac action has chiral symmetry, and cannot generate the $U(1)_A$-violating coupling. Such terms are present for non-vanishing quark masses and are proportional to the latter. This leaves us with the dimensional running and the self-interaction terms also present in the respective NJL-type four-quark model. The flow reads, see \cite{Gies:2003dp, Gies:2005as, Braun:2011pp, Mitter:2014wpa, Cyrol:2017ewj, Braun:2019aow, Eichhorn:2011pc}.
\begin{align}
\partial_t \bar\lambda_{\mathrm{top}} =  &\, \left(2 + \frac{17}{6\pi} g_N\right)\bar\lambda_{\mathrm{top}}+ \frac{(2N_c+3)N_c-1}{2N_c \pi^2 }\bar\lambda_{\mathrm{top}}^2 \nonumber \\
& \hspace{1.5cm}+ \f{N_c-1}{4 N_c\pi^2} \left (4 \bar\lambda_{\mathrm{top}} +\bar\lambda_q\right) \bar\lambda_q 
\,. \label{eq:TopFlow} 
\end{align}
The RG flows of $g_N(k)$ and $g_s(k)$ solving the flow equations \eqref{eq:TopFlow} and \eqref{eq:gNFlow} are shown by Fig.~\ref{fig:gs-gNFlow}.

In the NJL-type model, spontaneous chiral symmetry breaking is related to a divergence of the coupling $\lambda_q$. This is easily understood within the bosonised version of the model: There, the mass squared $m_\phi^2$ for the  composite bosonic field $\phi \sim (\bar{q} q)$ is related to the four-quark coupling, $\lambda_q \sim 1/m_\phi^2$. At $m^2_\phi<0$ the effective potential of the composite field develops non-trivial minima. Hence, at $m^2_\phi=0$, the theory goes from the symmetric into the broken phase. A quantitative evaluation of spontaneous chiral symmetry breaking requires more refined approaches. Such a refined analysis can be performed e.g.\  with the weak RG method~\cite{Aoki:2014ola, Aoki:2017rjl, Grossi:2019urj} or within dynamical bosonisation~\cite{Aoki:1999dw, Gies:2001nw, Gies:2002hq, Mitter:2014wpa, Braun:2014ata, Rennecke:2015eba, Cyrol:2017ewj, Fu:2019hdw, Floerchinger:2009uf, Denz:2019ogb}.

\section{Gravitational instanton}
\label{sec:GravInst}

\begin{figure}[tbp]
 \centering
 \includegraphics[width=0.45\textwidth]{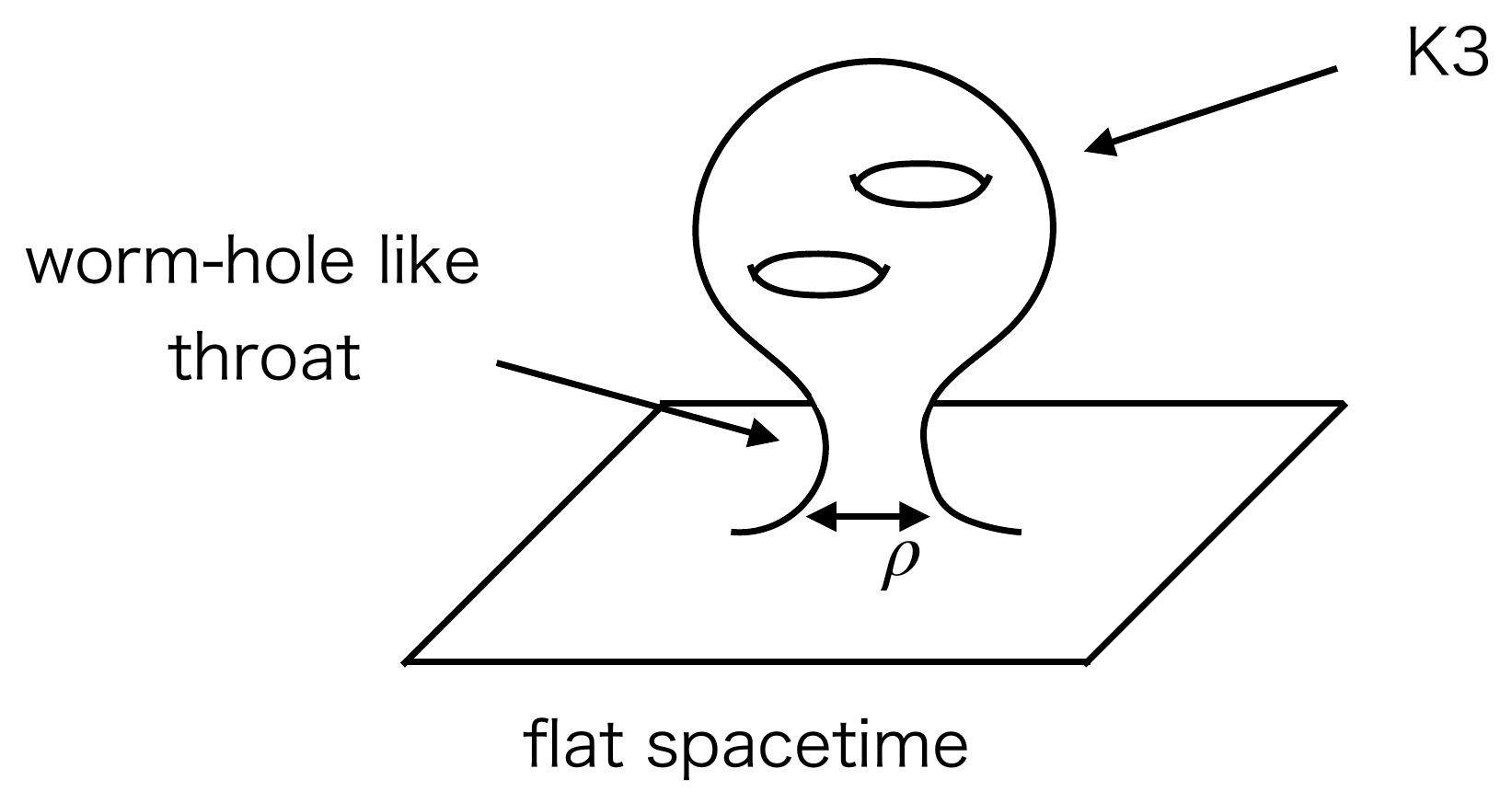}
 \caption{The K3-surface connected by a wormhole-like throat with the flat spacetime, as discussed in \cite{Hebecker:2019vyf}. The typical size of the wormhole is given by $\rho$.}
 \label{fig:GravInst}
\end{figure}

We here give a brief review on gravitational instantons, for more comprehensive reviews, see Refs.~\cite{Eguchi:1978gw,Eguchi:1980jx}. They are (anti-) selfdual solutions of the vacuum Einstein equations: $\tilde{R}_{\mu\nu}= \pm R_{\mu\nu}$ with $\tilde{R}_{\mu\nu} \equiv  \epsilon_{\mu\nu\lambda\rho}R^{\lambda\rho}$. Gravitational instantons are Ricci flat, $R_{\mu\nu}=0$, and provide spacetime manifolds with non-trivial topology, characterised by the signature $\tau$, 
\begin{align}\label{eq:tau}
  \tau[\mathcal{M}] 
  &= -\f{1}{96 \pi^2} \int \df^4x~ \sqrt{g} ~\epsilon^{ \mu\nu\lambda\rho} {R_{\mu\nu}}^{\alpha\beta} R_{\lambda\rho\alpha\beta}\,,
\end{align}
with the 2-form curvature $\bm{R}$, 
\begin{equation}
 {(\bm{R})_\mu}^\nu = \f{1}{2} {R_{\alpha\beta\mu}}^\nu \df x^\alpha \df x^\beta\,.
\end{equation}
The Dirac index $I[\mathcal{M}]$ is defined by the difference between the numbers of the positive and negative chirality eigenmodes of the Dirac operator $\s{D}$ on a background spacetime $\mathcal{M}$,
\begin{equation}
 I[\mathcal{M}] \equiv n_+ - n_-\,.
\end{equation}
The Atiyah-Patodi-Singer (APS) index theorem \cite{atiyah_patodi_singer_1,atiyah_patodi_singer_2,atiyah_patodi_singer_3} states that the analytic Dirac index is related to topological invariants,   
\begin{align}
  I[\mathcal{M}] = \f{1}{8} \tau [\mathcal{M}] + \f{1}{2} ~\eta_D[\partial \mathcal{M}]\,,\label{eq:APSindex}
\end{align}
where the signature $\tau$ has been given in \eq{eq:tau}, and $\eta_D[\partial \mathcal{M}]$ is the eta-invariant defined on the boundary $\partial\mathcal{M}$, for recent work see \cite{Fukaya:2017tsq,Fukaya:2019qlf}. The first term on the right-hand side of Eq.~\eqref{eq:APSindex} is a contribution from the bulk, while the second one is that from the boundary $\partial\mathcal{M}$. For manifolds without boundaries, the latter vanishes and thus the equation reduces to the Atiyah-Singer index theorem \cite{Atiyah:1963zz,Atiyah:1968}, 
\begin{equation}
 I[\mathcal{M}]=-\f{1}{768 \pi^2} \int \df^4x~ \sqrt{g} ~\epsilon^{ \mu\nu\lambda\rho} {R_{\mu\nu}}^{\alpha\beta} R_{\lambda\rho\alpha\beta}\,.
\end{equation} 
Manifolds $\cal M$ with non-vanishing index $[\mathcal{M}]$ sustain chiral zero-modes of the Dirac operator, that trigger anomalous chiral breaking symmetry breaking via the chiral anomaly. 

We now discuss three examples for gravitational instantons: the Eguchi-Hanson metric \cite{Eguchi:1979yx}, the Taub-NUT metric \cite{Hawking:1976jb,Eguchi:1977iu}  and the K3-surface. The first two metrics describe non-compact manifolds that approach locally flat Euclidean spaces for asymptotically large distances. They have the vanishing index $I[\mathcal{M}]$ because their boundary contribution $\f{1}{2} ~\eta_D[\partial \mathcal{M}]$ cancels the bulk contribution $\f{1}{8}\tau$ in Eq.~\eqref{eq:APSindex}. On the other hand, The K3-surface is the only closed and compact manifold that satisfies the self-dual condition. Unfortunately the explicit form of its metric is still unknown. It is known nevertheless that K3 surface has the non-zero index $I=2$, and thus the Dirac operator has two chiral zero-modes. Therefore, it is a promising candidate for a manifold inducing chiral-symmetry breaking effects. 

However, the K3-surface itself is not suitable to our argument because it is a compact manifold with a typical size of $1/M_\mathrm{pl}$. Hence it cannot be regarded as a localised object in our universe. An alternative choice is a manifold where the K3-surface and the flat Euclidean space $\mathbb{R}^4$ are connected by the wormhole-like throat (Fig.~\ref{fig:GravInst}), as introduced in \cite{Hebecker:2019vyf}. This manifold has the same index as the K3-surface, because its boundary is $S^3$, which has a vanishing eta invariant, $\eta_D[S^3]=0$ (one can always make the configuration compact by adding a single point at infinity). Note, that this whole spacetime is neither a self-dual manifold nor is it a solution of the vacuum Einstein equation due to the wormhole throat. Accordingly it is not a  saddle point of the classical action, which suggests that it is strongly suppressed in the path integral. In the next  Sec.~\ref{sec:tHooft} we will argue that this is not the case, and we can resort to semi-classical arguments.

\section{Topological contributions to the 't Hooft coupling}
\label{sec:tHooft}

The classical $U(1)_A$-symmetry of the massless Gravity-QCD system is broken by quantum effects induced by both, QCD and gravitational instantons. These breaking effects originate from the  integration of fermionic zero-modes localised around the instantons. In the present Section we derive the QCD and gravitational instanton contributions to the four-quark flows. While the derivation of the QCD-instanton contribution simply reminds on the derivation done in \cite{Pawlowski:1996ch}, the derivation of the latter is new, and is done in analogy of the QCD-case.

\subsection{Instanton-effects in the dilute gas approximation} \label{sec:Instantons+Beyond}
In the dilute gas approximation with localised and dilute instantons they leads within a semi-classical expansion to the 't~Hooft interaction in \eq{eq:Gferm} with the coupling $\lambda_{\mathrm{top}}$. In the semi-classical approximation for the present Gravity-QCD system the coupling receives additive contributions from QCD and gravitational instantons, 
\begin{align}
\lambda_{\textrm{top}} \simeq  \lambda_{\textrm{top}}^{(\textrm{glue})}+ \lambda_{\textrm{top}}^{(\textrm{grav})}\,.
\end{align}
The contribution from an QCD-instanton with size $\rho$ in the dilute gas approximation is given by \cite{tHooft:1976rip,tHooft:1976snw},  
\begin{equation}
\lambda_\mathrm{top}^{(\mathrm{glue})} \sim \rho ^{3 N_f-4} \exp \left(-\frac{8\pi^2}{g_s^2(1/\rho)}\right)\,. \label{eq:QCDInst}
\end{equation}
In the present fRG-approach with an infrared cutoff scale the QCD-instanton contributions  $\lambda_\mathrm{top}^{(\mathrm{glue})}$ have been studied in \cite{Pawlowski:1996ch} within an expansion about background instantons. In \cite{Pawlowski:1996ch} it has been shown that the semi-classical analysis including the relevance and effects of chiral zeromodes holds true in the presence of an infrared cutoff. The cutoff scale $k$ serves as an infrared cutoff for the instanton size with $\rho \lesssim 1/k$ and the flow integrates out instantons with the size $\rho\propto 1/k$. The latter fact is already reflected in the running coupling $g_s^2(1/\rho)$ evaluated at the momentum scale $1/\rho$. This leads us to the flow of the dimensionless four-quark coupling $\bar\lambda_\mathrm{top}^{(\mathrm{glue})}=k^2 \lambda_\mathrm{top}^{(\mathrm{glue})}$ for $N_f=2$ with 
\begin{equation}
\partial_t \bar \lambda_\mathrm{top}^{(\mathrm{glue})} \propto \partial_t\left[ e^{-\frac{8\pi^2}{g_s^2(k)}}\right] = \frac{8\pi^2\beta_{g_s^2}}{g_s^2(k)}\, e^{-\frac{8\pi^2}{g_s^2(k)}}\,, \label{eq:FlowQCDInst}
\end{equation}
with $\beta_{g_s^2}= \partial_t \log g_s^2$. This prefactor arises from the $t$-derivative of \eq{eq:QCDInst} with $\rho\propto 1/k$.  

We use these results to also estimate the magnitude of $\lambda_\mathrm{top}$ as well as $\partial_t \lambda_\mathrm{top}$ induced from gravitational instantons. In particular, we consider the spacetime consisting of the K3 surface and the flat spacetime connected by the wormhole, see Fig.~\ref{fig:GravInst}. For a given typical size $\rho$ of the wormhole $\rho$, we can estimate its effect based on the naive dimensional analysis as 
\begin{align}
\lambda_\mathrm{top}^{(\mathrm{grav})}\sim \rho ^{3N_f-4}\, e^{-S[\mathcal{M}]}\,.\label{eq:GravInst}
\end{align}
The classical action $S[\mathcal{M}]$ of this geometry is roughly given by 
\begin{align}
 S[\mathcal{M}] & \sim \Bigl(S_{\mr{flat}} + S_\mr{K3} + S_\mr{wormhole}\Bigr)\sim M_\mathrm{pl}^2 \rho^2 \sim \f{\rho^2}{G_N}\,,
\end{align}
where we have used that the Planck mass squared is the inverse Newton gravitational constant $M_\text{pl}^2=1/(8\pi G_N)$. Moreover, $S_\mr{flat}=S_\mr{K3} =0$, as the K3-surface is Ricci flat. 

In analogy to the QCD-analysis, the infrared cutoff term for gravity restricts the size of gravitational instantons to those with $\rho\lesssim 1/k$. The Newton constant $G_N$ should be identified with the running effective coupling constant at the scale $k$, $G_N(k)=g_N(k)/k^2$. Similarly to the contributions of QCD-instantons we now can derive the flow of the dimensionless coupling $\bar \lambda_\mathrm{top}^{(\mathrm{grav})} =k^2\lambda_\mathrm{top}^{(\mathrm{grav})} $ for $N_f=2$. The $t$-derivative hits the exponent in  \eqref{eq:GravInst} and we arrive at 
\begin{align}
\partial_t \bar \lambda_\mathrm{top}^{(\mathrm{grav})}   \propto \partial_t \left[ e^{- \f{1}{g_N(k)}}\right] =  \frac{\beta_{g_N}}{g_N(k)}e^{- \f{1}{g_N(k)}}\,. \label{eq:FlowGravInst} 
\end{align}
with $\beta_{g_N}=\partial_t \log g_N$. Not surprisingly, the instanton contributions both from QCD, \eqref{eq:FlowQCDInst}, and gravitational instantons, \eqref{eq:FlowGravInst}, have the same form. However, they differ qualitatively by the qualitatively different scale-dependence of the strong coupling and the Newton coupling: 

The QCD-coupling constant $g_s$ is asymptotically free, i.e., approaches to the Gau\ss ian fixed point $g_{s*}=0$ in the UV limit. Accordingly, the QCD-contributions  of small-size instantons are strongly suppressed while those from large-size ones are not suppressed due to the growing coupling. 

In turn, in asymptotically safe gravity-matter systems, the running Newton constant $g_N(k)$ approaches to a non-trivial fixed point $g_{N\ast} \neq 0$ above the Planck scale with $k/M_\mathrm{pl}\to\infty$. Consequently, the effects of small-size gravitational instantons with $\rho \ll M_\mathrm{pl}^{-1}$ ($\rho \sim 1/k $) are independent of $\rho$ and hence not suppressed. In the infrared with $k \ll M_\mathrm{pl}$, the dimensionless Newton constant $g_N(k)\sim k^2$. Accordingly, the contributions from gravitational instantons decay  exponentially below the Planck scale. In summary, the size of the gravitational instantons have an effective IR cutoff: $\rho\lesssim M_\mathrm{pl}^{-1}$.

\subsection{Flow of the 't Hooft coupling}\label{sec:floweq} 

The two estimates for the instanton contributions in the dilute-gas approximation to the flow of the dimensionless 't Hooft coupling, \eq{eq:FlowQCDInst} (QCD-instantons) and \eq{eq:FlowGravInst} (gravitational instantons) allows us to analyse gravitational catalysis of anomalous chiral symmetry breaking in the Gravity-QCD system. While the qualitative scale-dependences are also present in a fully non-perturbative setup, the non-perturbative quantitative determination of the prefactor is rather difficult: Firstly, due to its topological nature it is difficult (even though possible) to devise a reliable approximation to the full system, whose flow incorporates the generation of the topological flows, and in particular that of the  $1/g_N$-prefactor. This intricacy is already well-known from and studied in QCD-flows. Secondly, the dynamics of space-time may strengthen or weaken the topological effects. This -global- dynamics of space-time is an additional complication not present in QCD. 

\begin{figure}[tbp]
	\centering
	\includegraphics[width=0.45\textwidth]{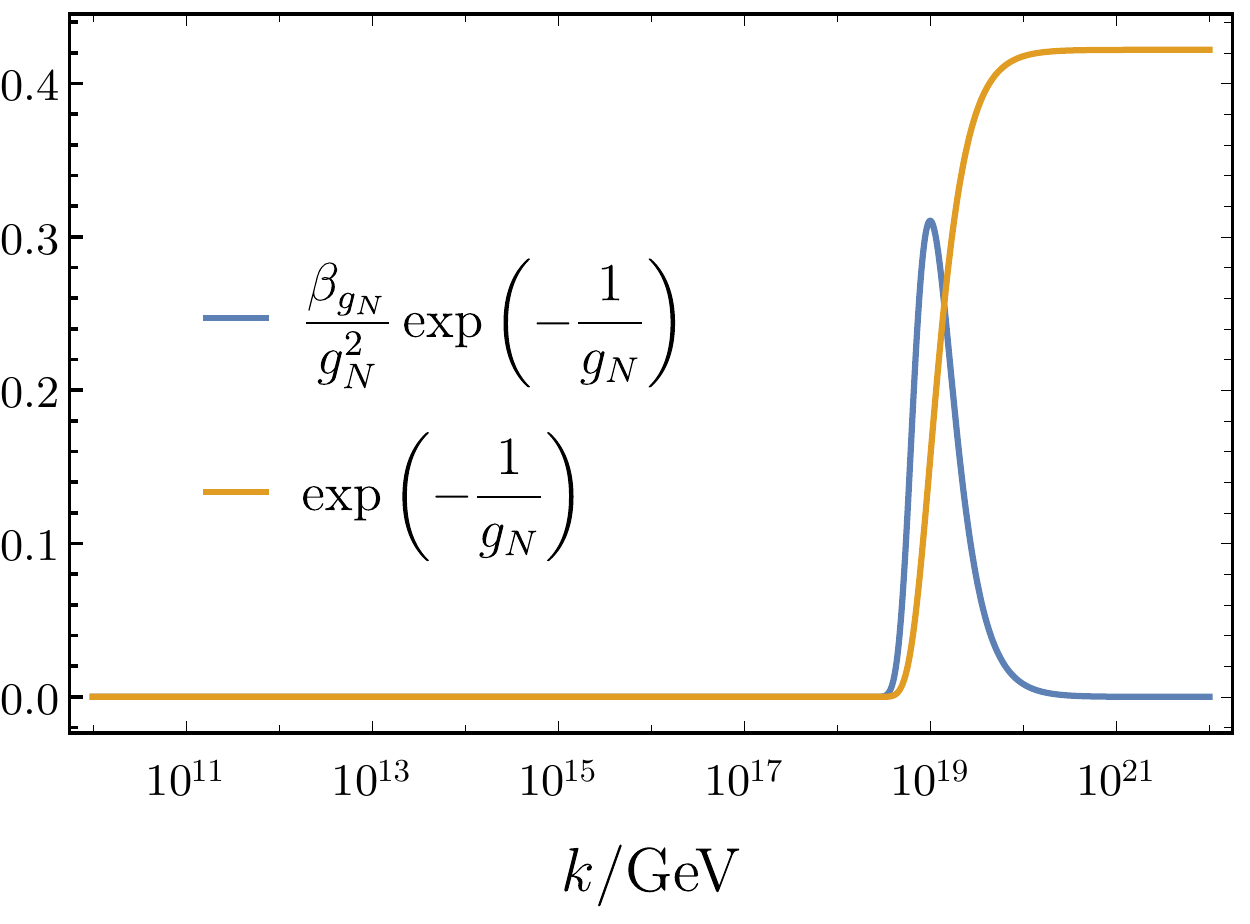}  
	\caption{Scale dependence of the topological terms in Eq.~\eqref{eq:TopFlowAll}. The Newton coupling $g_N(k)$ is obtained from the IR-UV flow in Fig.~\ref{fig:gs-gNFlow}, that is initiated at $k=10^2~\textrm{GeV}$ close to the electroweak scale, and the initial conditions for $g_N$ and $\alpha_s=g_s^2/(4 \pi)$ are the physical ones, $g_N= 6.71 \times 10^{-35 }$, and $\alpha_s=0.118$. The first term (blue line) generates the running of $g_N$ in the exponents $e^{-S}$ of the topological terms. It is proportional to $\beta_{g_N}$ and is strongly peaked at $k \sim M_\textrm{pl}$. The second term (yellow line) is related to the integrated first contribution, and originates from instanton--four-quark interactions. 
	}
	\label{fig:TopTermFlow}
	
\end{figure}

In combination this suggests a phenomenological approach, which allows to analyse the respective parameter ranges of the prefactors and its impact on spontaneous symmetry breaking. This is very similar to instanton-liquid considerations in QCD, for a review see \cite{Schafer:1996wv}. Hence, in the current work we will consider the instanton contributions \eq{eq:FlowQCDInst} and \eq{eq:GravInst} with free prefactors $\gamma^{(1)}_{\textrm{glue/grav}}$ that parameterise the topological dynamics. This leads us to the instanton contributions $\partial_t \bar\lambda_\mathrm{top}^{\textrm{(Inst)}}$ the flow of $\bar\lambda_\mathrm{top}$ with
\begin{align}\nonumber 
\partial_t \bar\lambda_\mathrm{top}^{\textrm{(Inst)}}= &\,  \left( \gamma^{(1)}_{\textrm{grav}}\, \frac{\beta_{g_N}}{g_N(k)} +\gamma^{(2)}_{\textrm{grav}} \bar\lambda_{\mathrm{top}}\right) \,e^{- \f{1}{g_N(k)}}  \\[1ex] 
& \hspace{-.4cm}
+ \left( \gamma^{(1)}_{\textrm{glue}}\, \frac{8\pi^2\beta_{g_s^2}}{g_s^2(k)} +\gamma^{(2)}_{\textrm{glue}} \bar\lambda_{\mathrm{top}}\right)  \, e^{-\frac{8\pi^2}{g_s^2(k)}}\,.
\label{eq:FlowInst} \end{align}
In \eq{eq:FlowInst}, the term proportional to the $\beta$-functions of the Newton coupling and the QCD coupling originates from the dynamics of the exponential factor $e^{-S}$ that depends on the running couplings. The terms proportional to $\bar\lambda_{\mathrm{top}}$ are generated from interactions between the instanton and $\bar\lambda_\text{top}$. They are absent for vanishing $U(1)_A$-breaking with $\bar\lambda_{\mathrm{top}}=0$. Eq.~\eq{eq:FlowInst} allows us to evaluate the phenomenological consequences of different scenarios by scanning though the  $\gamma^{(1)},\gamma^{(2)}$ parameter space: the latter parameter parameterise the dynamics of the topological sector of the theory, and in particular the non-trivial interactions of instantons. In turn, the exponential factors in \eq{eq:FlowInst} carry the fluctuation dynamics of QCD and gravity. As a showcase example we depict their scale-dependence in Fig.~\ref{fig:TopTermFlow} for the IR-UV flows already shown in Fig.~\ref{fig:gs-gNFlow}, the respective flow is initiated at $k=10^2$\,GeV close to the electroweak scale with the physical values of Newton coupling, $g_N= 6.71 \times 10^{-35 }$, and strong coupling, $\alpha_s=0.118$. 

%
%
In combination, the topological part of the flow and the flow in the absence of topological effects, \eq{eq:TopFlow}, provide the full flow equation of $\bar\lambda_{\mathrm{top}}$, 
\begin{align}
\partial_t \bar\lambda_\mathrm{top} =&\, \left(2 + \frac{17}{6\pi} g_N\right)\bar\lambda_{\mathrm{top}} + \partial_t \bar\lambda_\mathrm{top}^{\textrm{(Inst)}}  \nonumber \\[1ex]
&\hspace{-1cm}+ \frac{(2N_c+3)N_c-1}{2N_c \pi^2 }\bar\lambda_{\mathrm{top}}^2 + \f{N_c-1}{4 N_c\pi^2} \left (4 \bar\lambda_{\mathrm{top}} +\bar\lambda_q\right) \bar\lambda_q\,, 
\label{eq:TopFlowAll} \end{align}
with $\partial_t \bar\lambda_\mathrm{top}^{\textrm{(Inst)}} $ in \eq{eq:FlowInst}. We are now in the position to discuss the generation of chiral symmetry breaking at trans-Planckian scales. To begin with, the QCD-instanton contributions are negligible in this momentum regime: for $k\gtrsim M_\mathrm{pl}$ a conservative estimate gives $g_s(M_\mathrm{pl})\lesssim  1/2$. This leads to an exponential factor 
\begin{equation}
\exp\left(-\frac{8\pi}{g_s^2}\right)\sim e^{-100}\,,
\end{equation}
and the QCD-terms are negligible unless $\gamma^{(1)}_{\textrm{glue}}$ or $\gamma^{(2)}_{\textrm{glue}}\bar \lambda_\textrm{top}$ are of the order $e^{100}$. Another option for increasing the QCD contributions are finite quark masses. Then, chiral symmetry is explicitly broken and $\partial_t \bar \lambda_\textrm{top}$ receives contributions proportional to the quark mass. However, we have checked, that as long as the quark masses are far smaller than $M_\mathrm{pl}$,  the suppression of QCD-contributions beyond the Planck scale holds true. This leaves us with the parameter set $(\gamma^{(1)}_\textrm{grav}, \gamma^{(2)}_\textrm{grav})$, which controls the size of $U(1)_A$-breaking as well as the phase structure in the trans-Planckian regime.

\section{Anomalous gravitational catalysis of chiral symmetry breaking}
\label{sec:GravCat}

In this Section we evaluate the anomalous gravitational catalysis of chiral symmetry breaking in the Gravity-QCD system within the approximation detailed in the previous Sections (\autoref{sec:QCD-gravity}, \autoref{sec:GravInst}, \autoref{sec:tHooft}). This approximation led to the set of flow equations in \eqref{eq:gNFlow}, \eqref{eq:gsFlow}, \eqref{eq:qFlow},  and \eqref{eq:TopFlowAll}.

\begin{figure*}[tbp]
 \centering
 \includegraphics[width=0.48\textwidth]{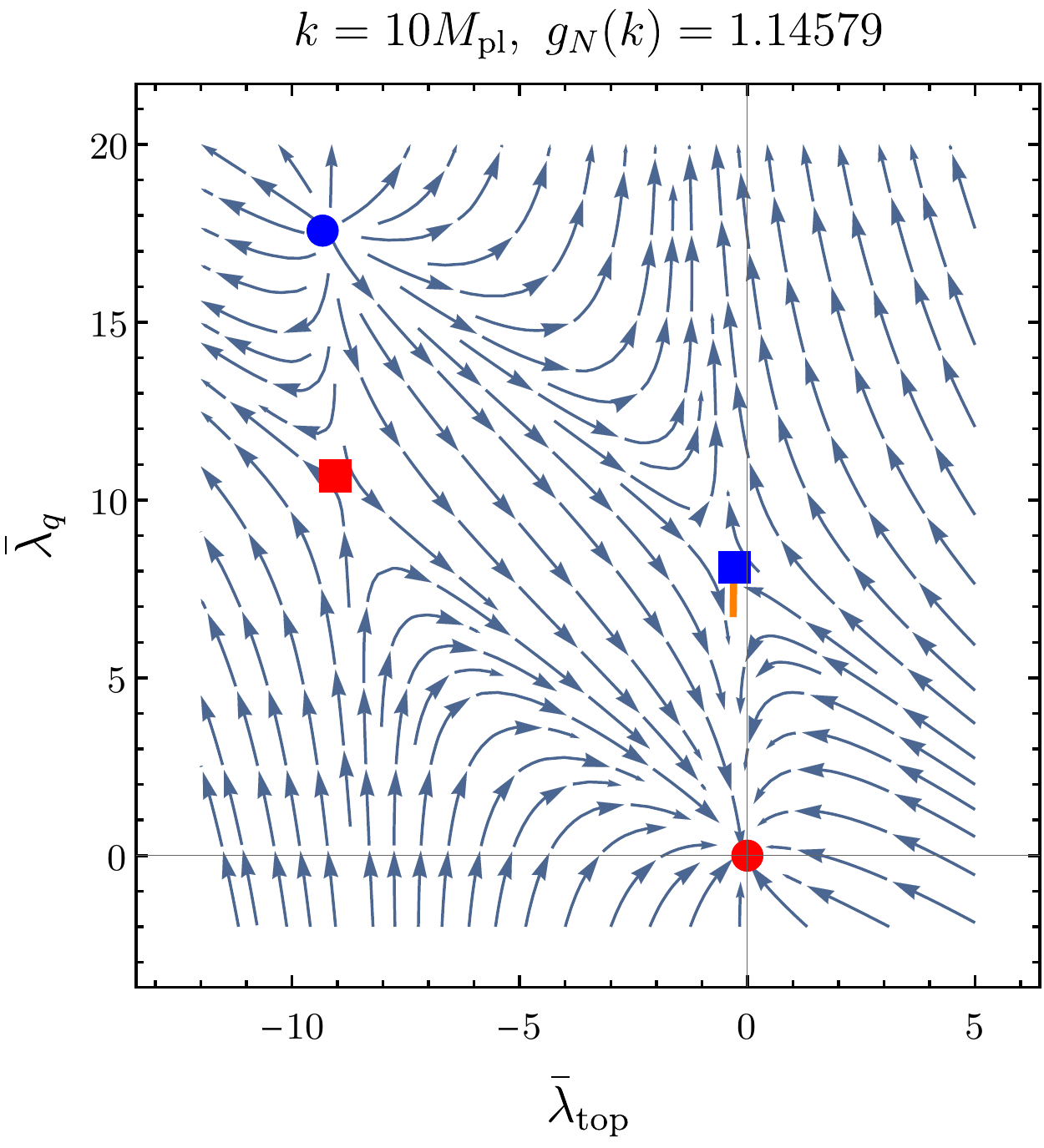}\hspace{1em}
 \includegraphics[width=0.48\textwidth]{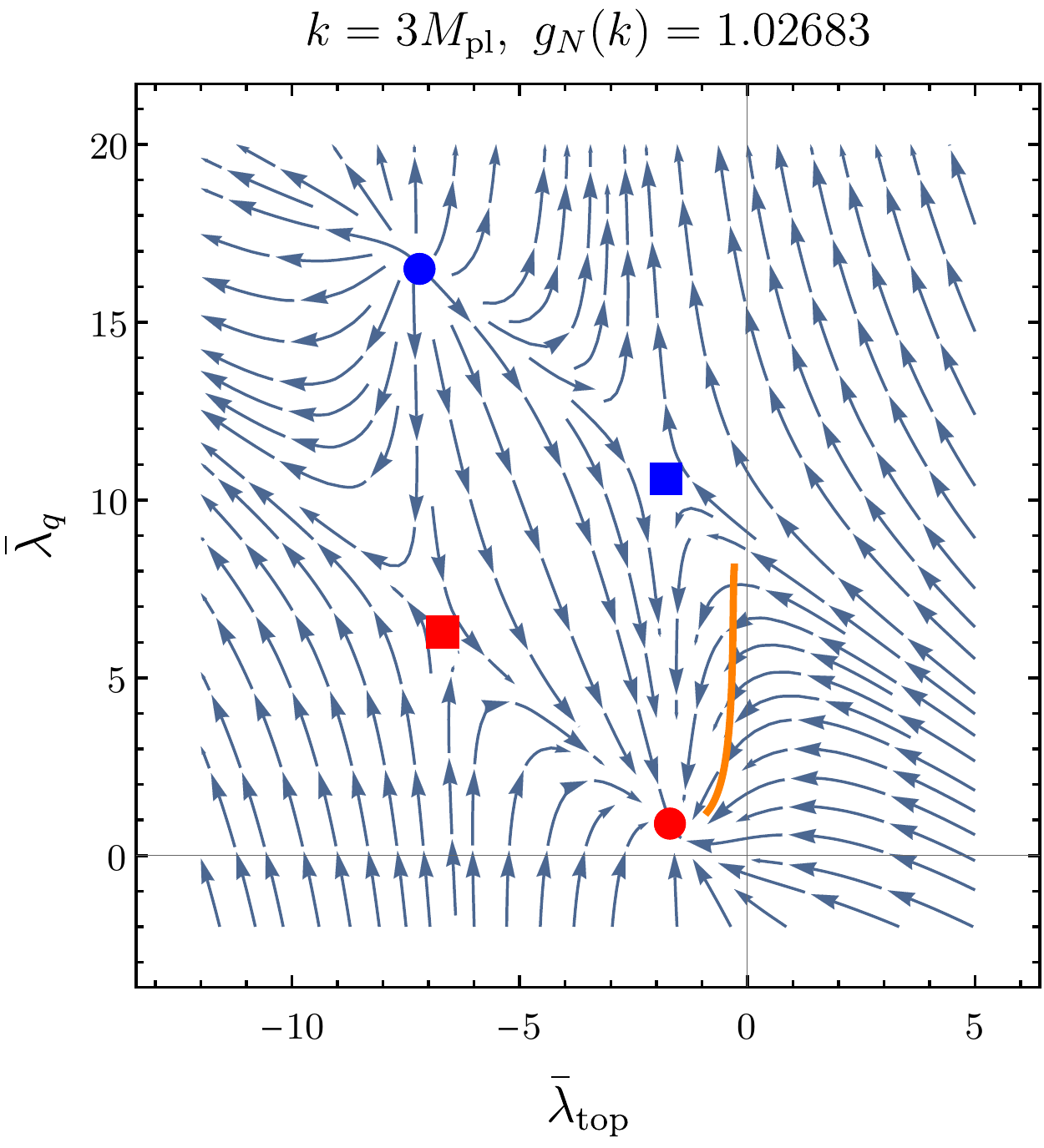} \\[5ex]
 \includegraphics[width=0.48\textwidth]{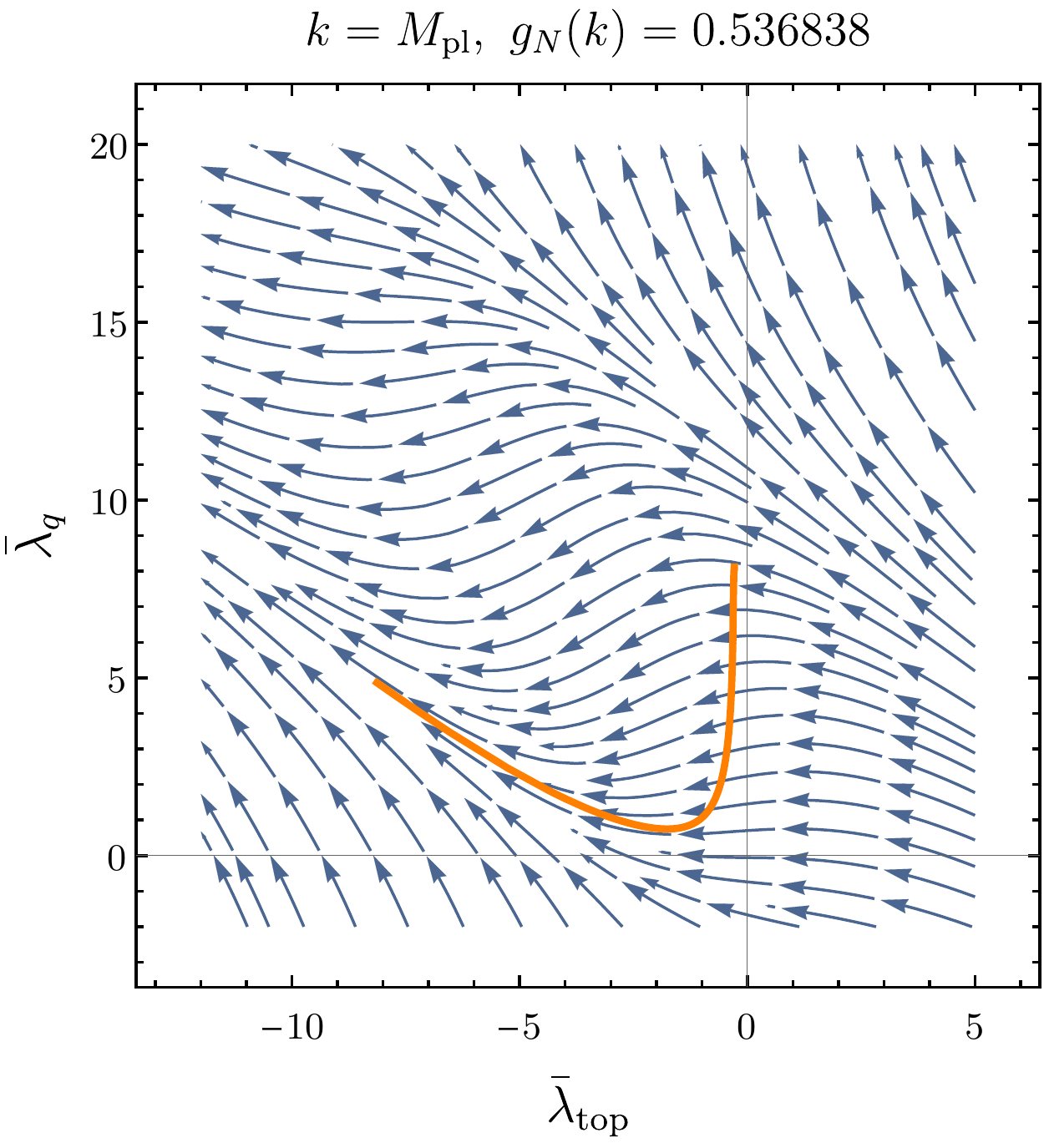}\hspace{1em}
 \includegraphics[width=0.48\textwidth]{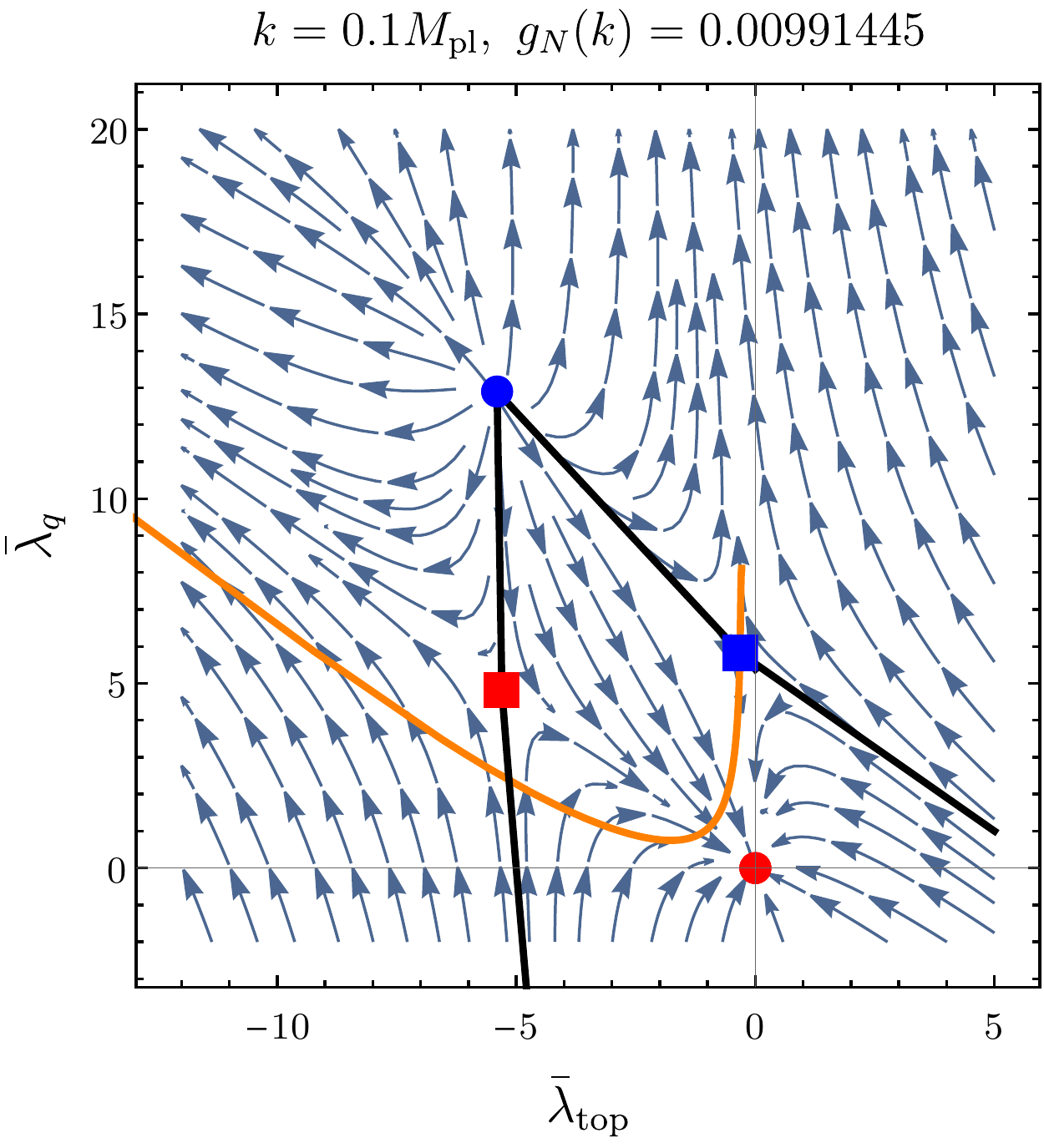}
 \caption{We show snapshots of the phase structure of the Gravity-QCD system given by \eqref{eq:gNFlow}, \eqref{eq:gsFlow}, \eqref{eq:qFlow},  and \eqref{eq:TopFlowAll}. The coefficients in the flow equation $\gamma^{(1)}_\textrm{grav}$, $\gamma^{(2)}_\textrm{grav}$ are taken as $(\gamma^{(1)}_\textrm{grav},\gamma^{(2)}_\textrm{grav})=(30,1)$.  Blue lines and arrows represent the RG flow for $\bar\lambda_\text{top}$ and $\bar\lambda_q$. For $k=10 M_\mathrm{pl}$, there are four fixed points. The (slightly shifted) Gau\ss ian fixed point is denoted by the red dot. The other fixed points are denoted by red square, blue dot and blue square. For $k\sim M_\mathrm{pl}$, the red dot and the red square collide and annihilate as do the blue dot and square. In this regime there are no IR-attractive fixed points. \\
 Orange line represents the UV-IR RG-trajectory of $(\bar\lambda_\text{top}(k), \bar\lambda_q(k))$. The flow is initiated close to the non-Gau\ss ian UV-fixed point with the initial condition \eq{eq:InCond} at $k=10^2 M_\textrm{pl}$. The initial values of the four-quark couplings in \eq{eq:InCond} are $(\bar\lambda_\text{top}, \bar\lambda_q)=(-0.2864,8.1135)$ and the flow is directed  towards the Gau\ss ian one for $k\gtrsim 3 M_\mathrm{pl} $. In the regime with $k\sim M_\mathrm{pl}$ The flow is pushed towards larger $\bar\lambda_q$ and $-\lambda_{\mathrm{top}}$ as a consequence of the fixed point annihilation, and runs into the regime with chiral symmetry breaking. In the present setup the scalar-pseudoscalar coupling $\bar\lambda_q$ diverges at $k\sim 0.1 M_\mathrm{pl}$.  
 }
 \label{fig:FlowStream}
\end{figure*}

Within this setup we evaluate spontaneous chiral symmetry breaking triggered by gravitational topological contributions, whose strength is parameterised by $\gamma^{(1)}_{\textrm{grav}}$ and $\gamma^{(2)}_{\textrm{grav}}$. Chiral symmetry breaking with the order parameter $\langle \bar q q \rangle \neq 0$, the chiral condensate, is tantamount to a divergence of the scalar-pseudoscalar coupling $\bar\lambda_q$ at a finite cutoff or momentum scale $k_\chi$. As the contribution of the gravitational-instanton term is strongly peaked at $k \sim M_\mathrm{pl}$, see Fig.~\ref{fig:TopTermFlow} the chiral symmetry breaking scale $k_\chi$ has to be proportional to the Planck mass. Consequently, we can restrict ourselves to this regime: $k\gtrsim M_\mathrm{pl}$. 

\subsection{Benchmark case: Setup}\label{sec:benchmark}
As a benchmark case we consider a set of parameters that triggers spontaneous chiral symmetry breaking rather generically, 
\begin{align}
	\label{eq:ben-ga}
	(\gamma^{(1)}_\textrm{grav}\,,\,\gamma^{(2)}_\textrm{grav})=(30\,,\,1)\,.
\end{align}
For this case we show constant $g_N$-slices of the phase structure in Fig.~\ref{fig:FlowStream}. The different $g_N$ are obtained for different cutoff scales $k\sim M_\textrm{pl}$: Blue lines and arrows in Fig.~\ref{fig:FlowStream} represent the RG flows for $\bar\lambda_\text{grav}$ and $\bar\lambda_q$, and the arrows indicates flows from the UV towards the IR. Then, 
gravity-induced chiral symmetry breaking is discussed at the example of a specific UV-IR trajectory, depicted  by the orange lines in the plots in Fig.~\ref{fig:FlowStream}: this RG-trajectory of $(\bar\lambda_\text{top}(k), \bar\lambda_q(k))$ is initiated 
in the vicinity of the non-Gau\ss ian fixed point (blue square). 
The full initial condition is given by 
\begin{align}
\label{eq:InCond}
(\bar\lambda_\text{top}, \bar\lambda_q , g_N, g_s) = (-0.2864,8.1135, 1.1589, 0.2722)\,,
\end{align}
at $k=10^2 M_\textrm{pl}$.
The respective UV-IR flows ends in the chiral-symmetry broken regime below the Planck scale.
 
Before we enter the discussion of the specific trajectory, we evaluate the general phase structure with the snapshots in Fig.~\ref{fig:FlowStream}: at large cutoff values, $k \gg M_\mathrm{pl}$, the topological terms are absent in the flow, and we encounter four fixed points (top-left panel). One is the (slightly shifted) Gau\ss ian fixed point, denoted by the red dot. The other, non-Gau\ss ian, fixed points are denoted by the red square, the blue dot and the blue square.  As the cutoff scale $k$ approaches the Planck mass $M_\mathrm{pl}$, the blue and red fixed point pairs approach (top-right panel), and finally merge and annihilate each other. For a cutoff scale regime about the Planck mass, $k\approx M_\mathrm{pl}$, this leaves us with a situation without  IR-attractive fixed points (bottom-left panel), as expected. In the infrared, for $\Lambda_\textrm{QCD} \ll k \ll M_\mathrm{pl}$, the fixed point structure is similar to that in the UV, since the gravitational topological contributions are decaying rapidly (bottom-right panel). In the deep infrared for scales $k\lesssim \Lambda_\textrm{QCD}$ we approach the regime of chiral symmetry breaking in QCD not discussed any further here.  

This pattern already allows for chiral symmetry breaking with an underlying mechanism, that is very similar but yet very different to that in spontaneous symmetry breaking in QCD triggered by the -intermediate- rise of the strong coupling, see e.g.~the reviews \cite{Braun:2011pp, Dupuis:2020fhh}: For this discussion we briefly recall the QCD situation by inspecting the flow of the scalar-pseudoscalar coupling \eq{eq:qFlow}. In QCD the gravity contributions are absent and for small gauge coupling $g_s$ the $\beta$-function $\beta_{\bar\lambda_q}=\partial_t \bar\lambda_q$ resembles that of an NJL-type model. This entails that for large enough coupling $\bar\lambda_q$ the flow is negative, which then triggers chiral symmetry breaking in the infrared. For rising gauge coupling $g_s$ the negative contribution from quark-gluon box diagrams, the term proportional to $g_s^4$ in \eq{eq:qFlow}, shifts $\beta_{\bar\lambda_q}=\partial_t \bar\lambda_q$ down, and beyond a critical coupling the $\beta$-function is negative for all $\lambda_q$, and chiral symmetry breaking is guaranteed. This shift is accompanied by  a reduction of the canonical running $2\bar\lambda_q$ by a term proportional to $-g_s^2 \bar\lambda_q$ that stems from the mixed quark-gluon exchange--$\bar\lambda_q$ diagram. This reduction of the canonical running supports the shift of the $\beta$-function, but it does not constitute the driving mechanism. In the deep infrared the quark-gluon exchange coupling drops again due to the QCD mass-gap, and the $\beta$-function $\beta_{\bar\lambda_q}$ returns to the NJL-type form. Consequently, the accumulated strength of chiral symmetry breaking comes from a rather subtle interplay between the rise of the strong coupling at low momenta and the dynamical generation of the QCD mass-gap, for a detailed discussion see  \cite{Dupuis:2020fhh}. 

Within gravity-induced chiral symmetry breaking in the Gravity-QCD system the r$\hat{\textrm{o}}$le of the gauge-coupling terms is not taken over by the analogue graviton contributions proportional to $g_N^2$ (box diagrams) and $g_N$ (quark-graviton--$\bar\lambda_q$ diagram). It has been shown in e.g.~\cite{Eichhorn:2011pc,Eichhorn:2011ec, Eichhorn:2016vvy, Meibohm:2016mkp, Eichhorn:2017eht}, that this potential QCD-type mechanism of gravity-induced chiral symmetry breaking does not work in flat space-times. 

Instead, the shift and deformation part is taken over by the last term in the right hand side of \eq{eq:qFlow}: the r$\hat{\textrm{o}}$le of the shift contribution (box diagrams) is taken over by the four-quark term (fish diagram) with two $U(1)_A$-violating couplings proportional to $\bar\lambda_{\mathrm{top}}^2$, and the deformation part is taken over by the mixed four-quark fish diagram proportional to $\bar\lambda_{\mathrm{top}}\, \bar\lambda_q$. In short, the r$\hat{\textrm{o}}$le of $g_s^2$ in QCD is taken over by $\bar\lambda_{\mathrm{top}}$. In both cases chiral symmetry breaking is triggered by the rise of the respective couplings. However, while the rise of the strong gauge coupling $g_s$ towards lower momentum scales is driven by the standard dynamics of the SU(3)-gauge theory, in the present system the rise of the $U(1)_A$-violating coupling $\bar\lambda_{\mathrm{top}}$ is dominantly triggered by gravitational instantons, see \eq{eq:TopFlowAll} with \eq{eq:FlowInst}. 

\begin{figure}[tbp]
	\centering
	\includegraphics[width=0.5\textwidth]{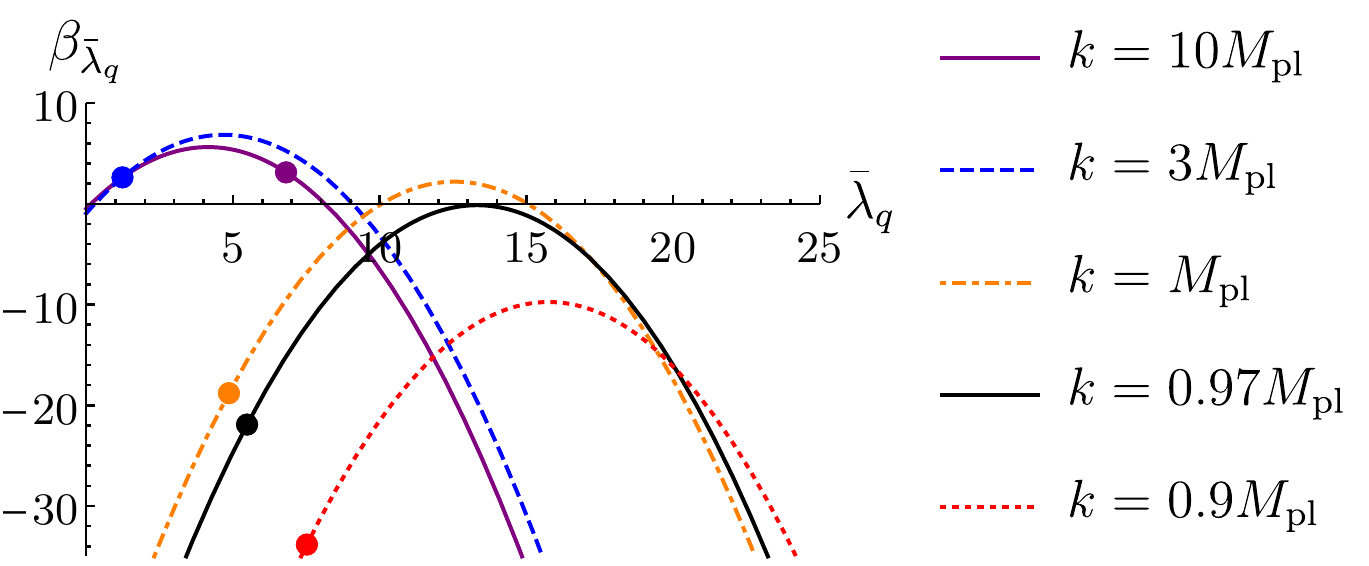}
	\caption{$\beta$-function $\bar{\lambda}_q$ of the scalar-pseudoscalar four-quark coupling $\bar\lambda_q$ for different cutoff values on the RG-trajectory (orange line) in Fig.~\ref{fig:FlowStream}. The values of the coupling and the $\beta$-function is signaled by respective blobs on the $\beta$-functions.\\ 
	Close to the UV-fixed point (violet straight line), the $\beta$-function resembles that of an NJL-type model: It has a finite UV-attractive fixed point $\lambda_q^*\neq 0$. It has a (close) Gau\ss ian infrared-attractive fixed point at vanishing coupling without chiral symmetry breaking, which is approached for initial couplings $\lambda_q^{(in)}<\lambda_q^*$. Finally, it features the chiral-symmetry breaking singularity, if the flow towards the infrared is initiated with $\lambda_q^{(in)}>\lambda_q^*$. \\		
		In the cutoff scale regime $k \sim M_\textrm{pl}$, the $\beta_{\bar{\lambda}_q}$ is shifted down by  the $U(1)_A$-violating coupling $\bar{\lambda}_\textrm{top}$ generated from the gravitational instantons (dashed blue line and dashed-dotted orange line). In the regime $0.5 \,M_\textrm{pl}\lesssim k \lesssim 0.97\, M_\textrm{pl}$, the $\beta$-function is negative for all $\bar{\lambda}_q$ and drives $\bar{\lambda}_q$ towards the chiral symmetry breaking singularity. 
		We have displayed the critical $\beta$-function at $k=0.97 \, M_{\textrm{pl}}$ (black straight line) and an exemplary one in this regime for $k=0.9 \, M_{\textrm{pl}}$ (dotted red line). \\ 
		For $k\ll M_\textrm{pl}$ the contributions from gravitational instantons decay exponentially, but the $\beta$-function stays in the negative regime due to the large $U(1)_A$-breaking coupling. Then, the $\beta$-function again resembles that of an NJL-type model, but it is not displayed here. }
	\label{fig: shift of beta function of lambdaq}
\end{figure}
This situation is summarised in Fig.~\ref{fig: shift of beta function of lambdaq}, the $\beta$-function $\beta_{\bar\lambda_q}$ shown there are snapshots along the RG-trajectory depicted by the orange lines in the single plots in Fig.~\ref{fig:FlowStream}: 

For small or vanishing $\bar\lambda_{\mathrm{top}}$ the $\beta$-function of the  scalar-pseudoscalar coupling $\bar\lambda_q$ resembles that of an NJL-type model. Moreover, as the initial four-quark couplings are below the critical values and in particular $\bar\lambda_q < \bar\lambda_q^*$ with $\beta_q(\bar\lambda^*_q)=0$, the four-quark couplings are driven towards the Gau\ss ian fixed point.

%
\begin{figure}[tbp]
	\centering
	\includegraphics[width=0.47\textwidth]{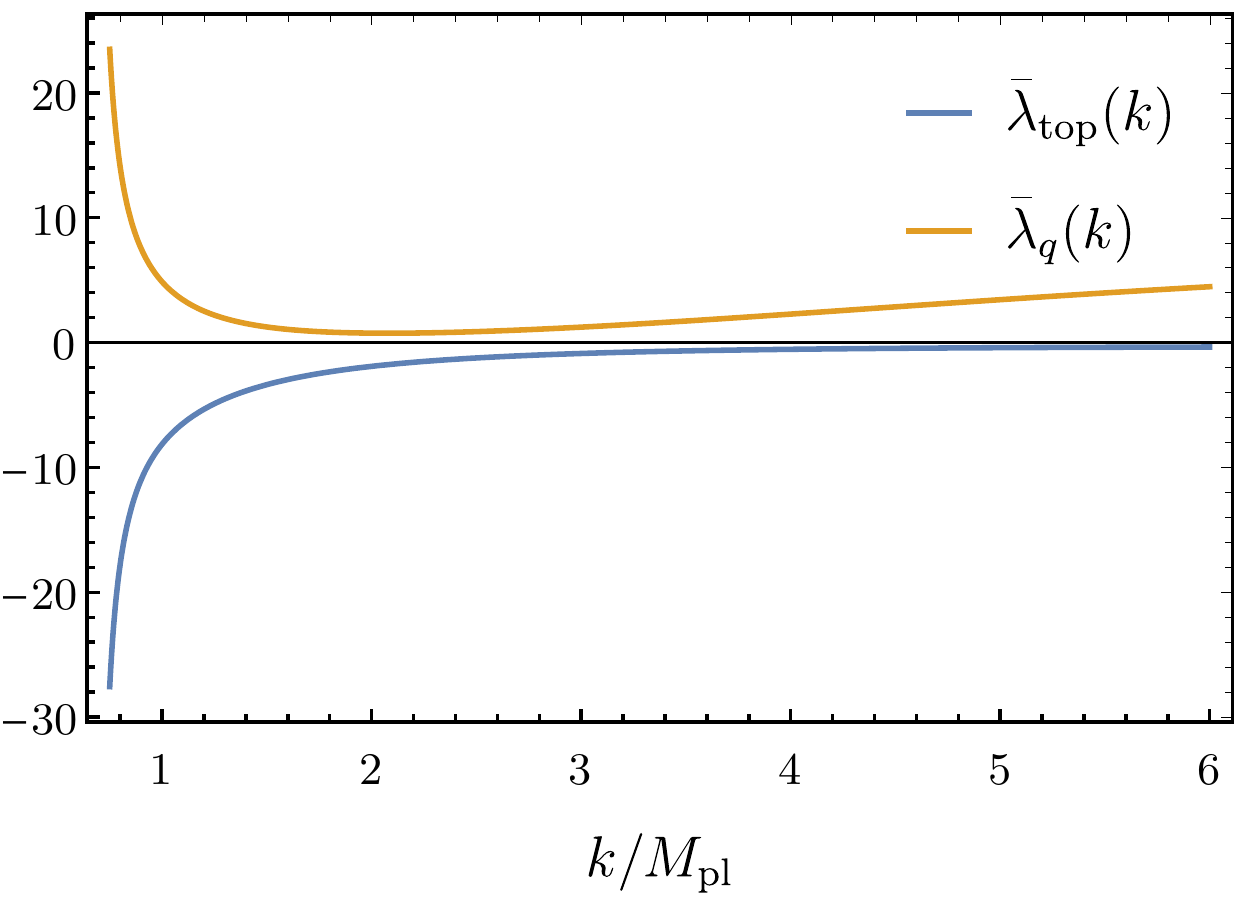}
	\caption{
		Scale-dependence of $\bar\lambda_q$ and $\bar\lambda_\text{top}$ for the UV-IR flow with the initial conditions \eq{eq:InCond} and $(\gamma^{(1)}_\textrm{grav},\gamma^{(2)}_\textrm{grav})=(30,1)$, the flow trajectory is the orange line in Fig.~\ref{fig:FlowStream}. Both four-quark couplings diverge at a finite momentum scale $k\sim 0.1 M_\mathrm{pl}$, which indicates that the chiral symmetry is spontaneously broken at a momentum scale much higher than the QCD scale $\Lambda_\mathrm{QCD}$ leading to Planck mass current quark masses.}
	\label{fig:LambdaFlow}
\end{figure}
\subsection{Benchmark case: Results}\label{sec:benchmarkResults}
 
 Now we discuss the specific trajectory in Fig.~\ref{fig:FlowStream}, as it evolved from the ultraviolet initial condition \eq{eq:InCond} to the infrared regime with spontaneous chiral symmetry breaking. We shall argue, that within our benchmark case with the parameters \eq{eq:ben-ga} the specific trajectory already covers generic initial conditions, in particular including those in the vicinity of the Gau\ss ian fixed point. The discussion will be carried out in terms of the single plots in Fig.~\ref{fig:FlowStream} and the respective $\beta$-functions in Fig.~\ref{fig: shift of beta function of lambdaq}:

\begin{figure*}[htbp]
	\centering
	\includegraphics[width=0.47\textwidth]{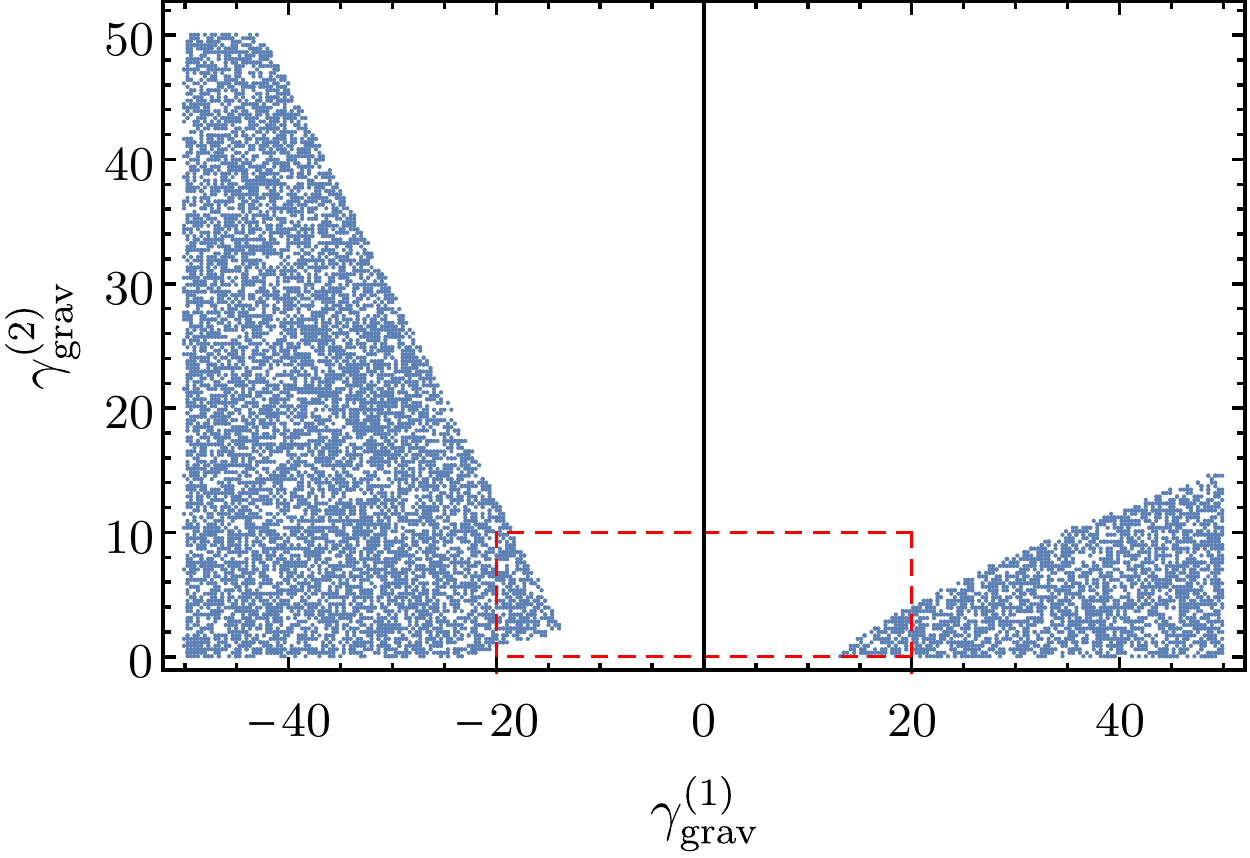}
	\hspace{0.03\textwidth}
	\includegraphics[width=0.47\textwidth]{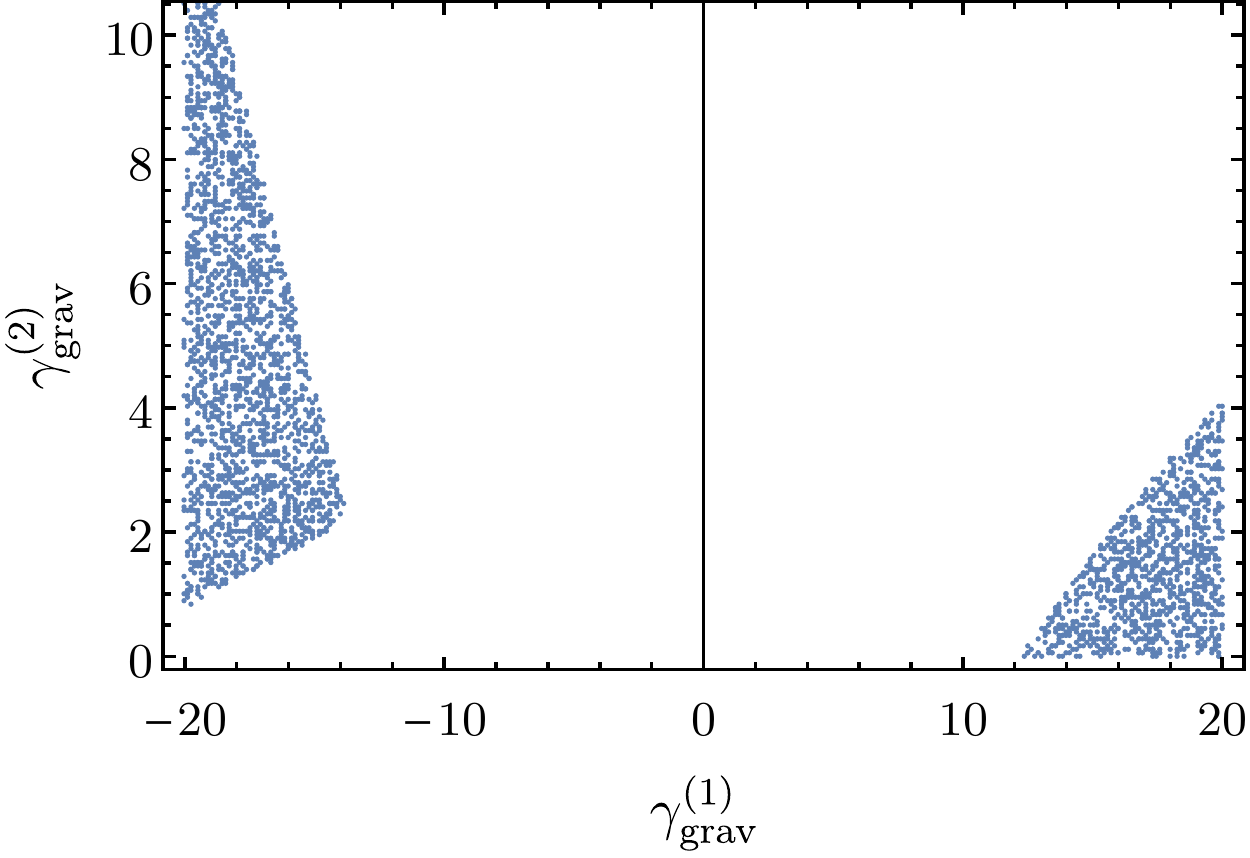}
	\caption{Spontaneous chiral symmetry breaking by anomalous gravitational catalysis, triggered by gravitational instantons in the $(\gamma^{(1)}_\textrm{grav},\gamma^{(2)}_\textrm{grav})$ plane. We have tested a large range of discrete parameter pairs for $-50<\gamma^{(1)}_\textrm{grav}<50$ and $0<\gamma^{(2)}_\textrm{grav}<50$ (left panel). Parameter pairs with chiral symmetry breaking are labelled with blue dots. We find that large values of $|\gamma^{(1)}_\textrm{grav}|$ favour chiral symmetry breaking. The right panel zooms into the region bounded by the red dashed lines in the left panel: $-20<\gamma^{(1)}_\textrm{grav}<20$ and $0<\gamma^{(2)}_\textrm{grav}<10$. In this regime anomalous gravitational catalysis does not take place (for $|\gamma^{(1)}_\textrm{grav}|\lesssim 14$). Hence, it is the physically viable regime. 
	}
	\label{fig:ParamSpace}
\end{figure*}

\begin{itemize}
\item[(i)]  $k= 10 \,M_\textrm{pl}$, \textit{Fig.~\ref{fig:FlowStream}, upper left panel}: For asymptotically large momentum (cutoff) scales $\bar\lambda_{\mathrm{top}}$ is rather small, both in the vicinity of the Gau\ss ian and the non-Gau\ss ian fixed point. Accordingly, the $\beta$-function $\bar\lambda_{\mathrm{top}}$ resembles that of an NJL-model, see Fig.~\ref{fig: shift of beta function of lambdaq} (straight (ultra-)violet line).

\item[(ii)] $k= 3\,M_\textrm{pl}$, \textit{Fig.~\ref{fig:FlowStream}, upper right panel}: Moving down towards the Planck scale regime, the contributions from gravitational instantons deform the fixed point structure and the blue-red pairs of fixed points moved towards each other. The topological contributions push down $\bar\lambda_{\mathrm{top}}$, see Fig.~\ref{fig: shift of beta function of lambdaq} (dashed blue line).  

\item[(iii)] $k= M_\textrm{pl}$, \textit{Fig.~\ref{fig:FlowStream}, lower left panel}:  For cutoff scales about the Planck scale, $k\propto M_\textrm{pl}$, the fixed points first annihilate, and then are absent for a short momentum range $0.5\, M_\textrm{pl}\lesssim k\lesssim 0.97\, M_\textrm{pl}$. In this regime the gravitational-instanton contributions trigger the rapid (negative) growth of the axial $U(1)_A$-violating coupling $\bar\lambda_{\mathrm{top}}$. Accordingly,  $\beta_{\bar\lambda_q}$ is shifted down and deformed. In Fig.~\ref{fig: shift of beta function of lambdaq} this is shown for $k=0.97 M_\textrm{pl}$ (dashed-dotted orange line) and $k=0.9 M_\textrm{pl}$ (dotted red line). Hence, in the cutoff scale regime $k\sim M_\textrm{pl}$ both couplings, $\bar\lambda_q$ and $ -\bar\lambda_{\mathrm{top}}$ grow large.

\item[(iv)] $k= 0.1\,M_\textrm{pl}$, \textit{Fig.~\ref{fig:FlowStream}, lower right panel}: Far below the Planck scale, $k\ll M_\textrm{pl}$, the contributions from gravitational instantons decay rapidly. Also, the standard (non-topological) gravity contributions decay and we are left with the QCD-$\beta$-function. However, in contrast to QCD the initial four-quark couplings, and in particular $\bar\lambda_{\mathrm{top}}$, are large. Accordingly, the $\beta$-function resembles the NJL-type $\beta$-function as for $k/M_\textrm{pl}\to\infty$, but  $\bar\lambda_q> \bar\lambda_q^*$, the UV-fixed point of the $\beta$-function. Hence, spontaneous chiral symmetry breaking is triggered: The scale-dependence of $\bar\lambda_\text{grav}$ and $\bar\lambda_q$ for the orange UV-IR trajectory in Fig.~\ref{fig:FlowStream} is shown by Fig.~\ref{fig:LambdaFlow}. The couplings $\bar\lambda_q$ and  consequently also $\bar\lambda_\text{grav}$ diverge at finite momentum scale $k_\chi\sim 0.1 M_\mathrm{pl}$. As stated before, this divergence is tantamount to chiral symmetry breaking, $\braket{\bar{q}q} \neq 0$. In the case discussed here, it is  induced by the topological contributions from gravitational instantons. We call this phenomenon {\it anomalous gravitational catalysis for chiral symmetry breaking}.
\end{itemize}

This closes our discussion of the generic case with spontaneous chiral symmetry breaking with initial conditions close to the non-Gau\ss ian fixed point. However, for $M_\textrm{pl} \lesssim k\lesssim 3 M_\textrm{pl}$ the coupling $\bar\lambda_{\mathrm{top}}$ is close to the Gau\ss ian fixed point. This entails that we may as well have started at the Gau\ss ian fixed point with our specific trajectory. This statement holds true generically for parameter pairs$(\gamma^{(1)}_\textrm{grav},\gamma^{(2)}_\textrm{grav})$ that lead to spontaneous chiral symmetry breaking, if initiated close the the non-Gau\ss ian fixed point. 
	
This leaves us with the following scenario: chiral symmetry breaking via anomalous gravitational catalysis takes place at about the Planck scale $M_\mathrm{pl}$. Hence, quarks acquire dynamical masses of the order of $M_\mathrm{pl}$, which  is at odds with the observed values. This allows us to put phenomenological constraints on the parameter space of $(\gamma^{(1)}_\textrm{grav},\gamma^{(2)}_\textrm{grav})$, the coefficients of anomalous gravitational catalysis for chiral symmetry breaking.

\subsection{Parameter range of anomalous spontaneous chiral symmetry breaking}\label{sec:GenResults}

With the analysis in the last section we can map-out the parameter regimes with and without anomalous gravitational catalysis of spontaneous chiral symmetry breaking. In Fig.~\ref{fig:ParamSpace} we show 
results for the parameter range for $-50 \leq \gamma^{(1)}_\textrm{grav} \leq 50$ and $ 0 \leq \gamma^{(2)}_\textrm{grav} \leq 50$: the blue dots represent the parameter set leading to the fixed point annihilation discussed in the benchmark case. Note in this context, that the fixed point annihilation is a necessary condition but not sufficient for the chiral symmetry breaking: the RG-flow can safely come back to the Gau\ss ian fixed point if the fixed point annihilation only holds true for a very short flow-time. However, this discrepancy between FP-annihilation and chiral symmetry breaking is only present within a very small parameter regime in the border between the blue-dotted area and the white one in Fig.~\ref{fig:ParamSpace}, and is insignificant in the present qualitative analysis. Thus, the blue-dotted area in Fig.~\ref{fig:ParamSpace} is the parameter set leading to the gravitational catalysis and hence this region excluded by experiment. 

In the vicinity of the boundary between the regimes with and without chiral symmetry breaking the generic analysis in the last section falls short. For parameter pairs in this boundary regime we expect, that anomalous gravitational catalysis shows some dependence on the choice of the initial condition. Then, a more quantitative analysis is required. Such a quantitative computation of the parameters $ (\gamma^{(1)}_\textrm{grav}, \gamma^{(2)}_\textrm{grav})$ in asymptotically safe gravity-matter systems is also mandatory for deriving phenomenologically viable constrains of the physically allowed area of UV-IR flows. This analysis and the evaluation of its phenomenological consequences is left to future work.

\section{Conclusions}
\label{sec:Discussion}
Whether asymptotically safe gravity-matter systems admit light fermions is a good probe for the observational validity of the theory. In the present work, we have investigated \textit{anomalous gravitational catalysis of chiral symmetry breaking}, triggered by gravitational instantons in asymptotically safe gravity: contributions from the latter topological configurations can deform the running of the four-quark interactions such, that anomalous spontaneous breaking of chiral symmetry is triggered at the Planck scale,  $k\sim M_\mathrm{pl}$. In this case, anomalous gravitational catalysis of chiral symmetry breaking generates quark or more generally fermion masses of the order of the Planck scale, which is at odds with the experimental observations. 

We have performed a phenomenological analysis reminiscent to instanton-liquid considerations in QCD, see \autoref{sec:floweq}: we have derived the flows of the Gravity-QCD system in the presence of  gravitational and QCD-instantons, see \eqref{eq:gNFlow}, \eqref{eq:gsFlow}, \eqref{eq:qFlow},  and \eqref{eq:TopFlowAll}. The prefactors of the gravitational topological contributions $(\gamma^{(1)}_\textrm{grav},\gamma^{(2)}_\textrm{grav})$ are taken as free parameters similar to that in the instanton-liquid in QCD. We have shown for an exemplary benchmark case, \eq{eq:ben-ga}, how  anomalous gravitational catalysis of chiral symmetry breaking occurs in the system. Snapshots of the phase structure of this case are found in Fig.~\ref{fig:FlowStream}, together with the respective snapshots of the $\beta$-function of the scalar-pseudoscalar coupling $\bar{\lambda}_q$ in Fig.~\ref{fig: shift of beta function of lambdaq}, and the scale-dependence of the couplings $(\bar{\lambda}_q, \bar{\lambda}_\textrm{top})$ in Fig.~\ref{fig:LambdaFlow}. A detailed discussion is provided in \autoref{sec:benchmarkResults}. 

In summary this allowed us us to determine the part of the parameter space in which spontaneous chiral symmetry breaking via anomalous gravitational catalysis takes place. We have found, that this effects is triggered in a  quite large regime of the parameter space, resulting in heavy (Planck-mass) fermions, see Fig.~\ref{fig:ParamSpace}: The blue-dotted regime signals chiral symmetry breaking via anomalous gravitational catalysis. This regime is excluded by experimental observations. 

The present work constitutes a first step towards a full quantitative analysis of gravitational anomalous chiral symmetry breaking within asymptotically safe gravity-matter systems. This quest for quantitative precision necessitates either bosonisation, e.g. \cite{Aoki:1999dw}, or dynamical hadronisation techniques, e.g.  ~\cite{Gies:2001nw, Gies:2002hq, Pawlowski:2005xe, Floerchinger:2009uf, Mitter:2014wpa, Braun:2014ata, Rennecke:2015eba, Cyrol:2017ewj, Jakovac:2018dkp, Fu:2019hdw, Denz:2019ogb}. Moreover, for a reliable grip on the symmetry-breaking pattern, the potential competing order effects as well as covering the large orders of magnitudes we also have to employ advanced numerical techniques such as the weak-RG~\cite{Aoki:2014ola,Aoki:2017rjl}, pseudospectral techniques, \cite{Borchardt:2015rxa, Borchardt:2016pif} or the discontinuous Galerkin methods \cite{Grossi:2019urj}. 

In terms of physics the first extension concerns the determination of the flows beyond the present instanton-liquid--type approximation. This concerns both the diagrammatic determination of the pair $(\gamma^{(1)}_\textrm{grav}, \gamma^{(2)}_\textrm{grav})$ as well as the global dynamics of space-time. 

Moreover,  the quantitative determination of the dynamically generated fermion masses and chiral condensate requires an extension of the approximation beyond the present four-fermi flows. in particular, for small values of $(\gamma^{(1)}_\textrm{grav}, \gamma^{(2)}_\textrm{grav})$, chiral symmetry breaking induced by anomalous gravitational catalysis may be postponed to lower momentum scales. Then, anomalous chiral symmetry breaking induced by the non-trivial topology  in QCD of the Yang-Mills gauge field may still show the imprint of a small relic from gravitational instanton effects. This could affect the dynamical quark masses and the chiral condensate, which may be tractable by precision high energy accelerator measurements.

\subsection*{Acknowledgements}
We thank Jens Braun, Holger Gies, Arthur Hebecker and Manuel Reichert for discussions. Y.\,H. thanks the Institut f\"ur Theoretische Physik, Universit\"at Heidelberg for the kind hospitality during his visit in which this work was initiated. Y.\,H. also thanks Masafumi Kurachi for his helpful support. This work is supported by EMMI, and the DFG Collaborative Research Centre SFB 1225 (ISOQUANT) as well as by the DFG under Germany's Excellence Strategy EXC - 2181/1 - 390900948 (the Heidelberg Excellence Cluster STRUCTURES). The work of Y.\,H. is partially supported by JSPS Grant-in-Aid for Scientific Research (KAKENHI Grant No. JP18J22733 and No. JP18K03655). The work of M.\,Y. is supported by the Alexander von Humboldt Foundation.

\bibliographystyle{apsrev4-1}
\bibliography{./references}

\end{document}